\documentclass[11pt,preprint]{aastex}

\newcommand\be{\begin{equation}}
\newcommand\en{\end{equation}}

\shorttitle{Transition Disks in CrA}
\shortauthors{Sicilia-Aguilar et al.}

\begin{document}

\title{Very Low-Mass Objects in the Coronet Cluster: The Realm of the Transition Disks\altaffilmark{1}}

\author{Aurora Sicilia-Aguilar\altaffilmark{2}, Thomas Henning\altaffilmark{2}, Attila Juh\'{a}sz\altaffilmark{2},}
\author{Jeroen Bouwman\altaffilmark{2}, Gordon Garmire\altaffilmark{3}, Audrey Garmire\altaffilmark{3}}

\altaffiltext{1}{Based on ESO program 079.C-0235}
\altaffiltext{2}{Max-Planck-Institut f\"{u}r Astronomie, K\"{o}nigstuhl 17, 69117 Heidelberg, Germany}
\altaffiltext{3}{The Pennsylvania State University, University Park, Pennsylvania 16802-6305, USA}

\email{sicilia@mpia.de}

\begin{abstract}

We present optical and IR spectra of a set of low-mass stars and brown
dwarfs in the Coronet cluster (aged $\sim$1 Myr), obtained with the multifiber 
spectrograph FLAMES on the VLT and with the IRS instrument on Spitzer. 
Most of the objects had been selected via their X-ray emission in a deep 
Chandra survey. The optical spectra reveal spectral types between 
M1 and M7.5, confirm the youth of the objects (via Li 6708\,\AA\ absorption), 
and show the presence of accretion (via H$\alpha$) and shocks (via forbidden 
line emission). The IRS spectra, together with IR photometry from the IRAC/MIPS
instruments on Spitzer and 2MASS, confirm the presence of IR excesses 
characteristic of disks around $\sim$70\% of the objects. Half of the disks do not 
exhibit any silicate emission, or present flat features characteristic of large 
grains ($>$6\,$\mu$m). The rest of the disks show silicate emission with 
indications of amorphous and crystalline silicate grains a few microns in size.
About 50\% of the objects with disks do not show near-IR excess emission, corresponding 
to the presence of ``transitional'' disks, according to their classical definition. 
This is a very high fraction for such a young cluster. The large number of ``transitional''
disks suggests lifetimes comparable to the lifetimes of typical optically thick disks.
Therefore, these disks may not be in a \textit{short-lived} phase, intermediate between Class II 
and Class III objects. The median spectral energy distribution of the disks in the Coronet
cluster is also closer to a flat disk than observed for the disks around solar-type stars
in regions with similar age.  The differences in the disk morphology and evolution in the 
Coronet cluster could be related to fact that these objects have very late 
spectral types compared to the solar-type stars in other cluster studies. Finally, the 
optical spectroscopy reveals that one of the X-ray sources is produced by a Herbig Haro 
object in the cloud.

\end{abstract}


\section{Introduction \label{intro}}

The formation of very low-mass stars and brown dwarfs (BD) has been a matter of 
discussion for a long time (Luhman et al. 2006). The difficulties in fragmenting a cloud or
stopping the envelope mass accretion at or near the hydrogen burning limit suggested 
that very low-mass objects may not form like solar-type stars, but require other 
mechanisms like ejection in multiple systems (Bate et al. 2005; Umbreit et al. 2005)
or photoerosion by nearby high-mass stars that can remove part of the initial core
material (Withworth \& Zinnecker 2004). At the same time, multiple studies have focused  
on the detection of accretion disks around very low-mass young stars and BD
(Fern\'{a}ndez \& Comer\'{o}n 2001; Natta \& Testi 2001; Jayawardhana et al. 2002, 2003;
Pascucci et al. 2003; Klein et al. 2003; Luhman et al. 2005; Mohanty 2004, 2005). 
H$\alpha$ spectroscopy reveals 
active accretion, and the excesses in the near- and mid-IR are consistent with 
disks similar to the disks around classical T Tauri stars (CTTS; Muench et al. 2001; 
Apai et al. 2002, 2005; Pascucci et al. 2003; Mohanty et al. 2004;
Jayawardhana et al. 2005; Scholz \& Jayawardhana 2008). IR spectroscopy finds further
similarities: dust evolution, grain growth/settling and disk flattening, formation of 
crystalline silicates, and inner holes (Furlan et al. 2005;  Apai et al. 2005; 
Gizis et al. 2005; Sterzik et al. 2004; Sargent et al. 2006; Natta et al. 2007;
Morrow et al. 2008; Henning 2008), which are thought to be the first steps in the 
formation of planetary systems, occur in these disks. Therefore, objects of all
masses may be formed by cloud fragmentation and subsequent disk formation and
accretion (Luhman et al. 2007).

Nevertheless, our understanding of disk evolution and accretion mechanisms in very 
low-mass stars and BD is still far from complete. The timescales for very low-mass 
disk evolution, expected to be longer than those of solar-type T Tauri stars (TTS) 
and Herbig Ae/Be stars (HAeBe; Sterzik et al. 2004; Dahm \& Hillenbrand 2007;
Bouy et al. 2007), are still very uncertain (Lada et al. 2006), due to the lack of 
statistically significant samples of coeval very low-mass stars and BD. IR observations 
of M-type stars reveal differences in innermost disk evolution (Kessler-Silacci et al. 2007; 
Sicilia-Aguilar et al. 2007), which may be the result of differences in the disk 
structure, presence of dead zones, and accretion mechanisms (Hartmann et al. 2006) and 
could affect the formation of planetary systems in such disks.
As it happens with the solar-type TTS, deeper and more complete multiwavelength
observations suggest that there is an important fraction of harder-to-detect,
flattened and transitional disks with inner holes around the lower-mass
objects. For these very low-mass objects, with masses $<$0.2 M$_\odot$ down to
the BD regime, the morphology of the disks could be different (Morrow et al. 2008),
but the difficulties in observing faint objects and determining the low-mass 
initial mass function in the regions translate into very small samples and
large uncertainties. If, like we observe in solar-type TTS, individual differences 
in disk evolution are remarkable even for objects of similar ages and spectral types, 
the knowledge of disks and accretion around very low-mass stars and BD and their 
evolution is highly limited at the present time, and may be biased toward ``special'' 
systems (with large IR excess and strong accretion rates). 

The Coronet cluster (also known as CrA) is an obscured star-forming region, 
located at a distance of $\sim$150 pc (d$\sim$ 130 pc according to Marraco \& Rydgren 1981 
and De Zeeuw et al. 1999; d$\sim$ 170 pc according to Knude \& Hog 1998), 
associated with the Herbig Ae star R CrA (Taylor \& Storey 1984) and a dense molecular 
core (Loren 1979). The Coronet cluster is relatively isolated at the edge of the Gould Belt. 
It is located out of the galactic plane, and it is probably part of the Sco OB2 Association 
(De Zeeuw et al. 1999) and of the large Sco-Cen region (De Geus et al. 1992). 
Cambr\'{e}sy (1999) estimated a total mass of 1600 M$_\odot$ 
for the whole cloud, using a distance of 170 pc. Casey et al. (1998) obtained an 
age of 3 Myr for the TY CrA multiple system, considering a distance of 130 pc; 
the cluster would be younger if the distance is larger. An age around 1 Myr is more 
consistent with the presence of nebular material and Class 0 objects
(Chen et al. 1997). Nisini et al. (2005) estimated an age of 0.5-1 Myr
based on 4 embedded objects, being therefore one of the closest and youngest
clusters. Submillimeter, millimeter, IR, and optical surveys (Henning et al. 1994; 
de Muizon et al. 1980; Wilking et al. 1985; Marraco \& Rydgren 1981) found a few HAeBe 
and TTS, suggesting a modest stellar population. The most massive stars in the region are 
the Herbig Ae star R CrA and the B9 star TY CrA. X-ray surveys using Einstein, XMM-Newton, 
ROSAT, and Chandra  (Walter 1986; Walter et al. 1997; Neuh\"{a}user et al. 1997, 2000; 
Hamaguchi et al. 2005a,b) identified several low-mass members. Deeper X-ray 
observations (Garmire \& Garmire 2003) found 118 potential (1$\sigma$) sources, many 
of which were consistent with young stars, some of them deeply embedded, which 
have not been described in detail yet. 

Here we present the results of optical and IR spectroscopy on
a collection of low- and very low-mass members of the Coronet
Cluster. The new and archive data, including the Chandra survey, optical 
spectroscopy with FLAMES/VLT, IR spectroscopy with IRS/Spitzer, and IRAC
and MIPS/Spitzer photometry are introduced in Section \ref{data}, where we 
also describe the selection methods of the sample of very low-mass objects.
The derived stellar and disk properties are presented in Section \ref{analysis},
and their implications for star formation and disk evolution are discussed in
Section \ref{discussion}. Finally, the results are summarized
in Section \ref{conclusions}.

\section{Selection of the sample, observations, and data reduction \label{data}}

Here we present the different observations (see Table \ref{obs-table} for
a summary). The sample was selected starting from a deep X-ray survey 
(Section \ref{data-xray}) and using additional optical candidates. 
The membership and spectral types were confirmed with the optical multifiber 
spectrograph FLAMES/VLT (Section \ref{data-opt}). Finally, the IRS instrument 
on Spitzer (Section \ref{data-irs}) was used to study the presence of disks and 
their characteristics, and the survey is completed with data from 2MASS and
IRAC/MIPS archival data (Section \ref{data-archive}). The Chandra field, together
with the optical candidates in the region, are completely covered in at least 
two of the IRAC channels (all the four channels in 95\% of the cases) and the 
MIPS maps.

\subsection{A very low-mass cluster in the CrA region: X-ray survey\label{data-xray}}

The starting point of this work were the X-ray detections of potential
low-mass objects in the region. The X-ray data were obtained on 2000 October 
7 with the Chandra X-ray Observatory (CXO).  The observation was made using the ACIS-I 
(AXAF CCD Imaging Spectrometer) array of CCDs in the timed-exposure mode 
(Garmire et al. 2003). This survey covered a 16.8'$\times$16.8' field centered on 
19:01:47.9 -36:58:22 J2000 with a 20 ks exposure. The data were aspect corrected 
using 10 stars in the same field with an estimated astrometric uncertainty of 0.5" 
for stars within 5 arc minutes of the aim-point. Source detection and location were 
determined using the WAVDETECT software (Freeman et al. 2002).  The energy range was 
limited to the range 0.5 - 8.0 keV to maximize the S/N.  The data were corrected   
for Charge Transfer Inefficiency (CTI) caused by radiation damage of the CCDs during
the first weeks after launch, which improves the energy resolution and charge
loss due to changes in the pattern of charge generated by X-ray interaction 
in the CCD pixels (Townsley et al. 2002). Flaring pixels were removed using the 
status filter and by examining the arrival time of each event.  A small subset of events 
was found to occur in the next exposure for  a given pixel and were missed by the status 
filter. These events were removed from the data as flaring pixel events. The X-ray 
properties of the confirmed members of the Coronet cluster are compiled in 
Table \ref{xray-table}. A larger list of sources and more detailed analysis will be 
reported in a future publication.

All the sources in Table \ref{obs-table} (except G-100, G-133, and G-136) were 
detected with more than 84\% confidence level. The limiting sensitivity for the 
observations was 2.1$\times$10$^{-15}$ erg cm$^{-2}$ s$^{-1}$ in the 0.5-8.0 keV band, 
which corresponds to a luminosity of 6$\times$10$^{27}$ erg s$^{-1}$. The depth of 
the survey ensures complete detection of all Class III and Class II objects 
down to the BD regime within the Chandra field (considering the typical relations
between bolometric and X-ray luminosities; Flaccomio et al. 2006), and an 
important number of Class I embedded sources. Some of the most extincted sources
(A$_V \geq$15 mag) may be not detected, but the high extinction region is
only a minimal part ($\sim$6\%) of the total Chandra field (Henning et al. 1994),
so most of the extincted sources are probably Class I/Class 0 objects with thick envelopes.
Due to the depth of the exposure and by comparison with other Chandra surveys, 
a fraction of the sources could be extragalactic (Garmire \& Garmire 2003). In 
addition, shocks and Herbig-Haro (HH) objects are known to produce X-ray emission 
(Pravdo et al. 2001; Favata et al. 2002; Raga et al. 2002), and 
several shocks have been identified in the optical
in the Coronet cluster (Graham 1993; Wang et al. 2004). Therefore, optical and IR 
data are essential in order to determine the nature of these sources.

\subsection{Optical spectroscopy \label{data-opt}}

In order to confirm the cluster membership and to obtain accurate spectral types,
we targeted the Coronet cluster with the multi-object optical spectrograph 
FLAMES on the UT2/VLT. Observations took place between April and June 2007 
within our program 079.C-0235. FLAMES, combined with the multifiber 
spectrograph GIRAFFE and the MEDUSA fibers, permits to obtain up to 130 spectra, 
including objects and sky positions. We observed one field centered on 
19:01:40 -36:56:30 J2000 with three different gratings (L682.2, L773.4, L881.7), 
providing intermediate resolution (R=5600-8600) spectra from 6440\,\AA\ to 9400\,\AA. 
This spectral range includes the main accretion feature (H$\alpha$ emission at 
6563\,\AA), Lithium I absorption line (signature of youth, 6708\,\AA), [S II] 
emission lines characteristic of shocks and Herbig Haro (HH) objects 
($\lambda\lambda$ 6716, 6731\,\AA) and a large number of photospheric features 
required for spectral type classification of late-type stars. 

The spectrograph fibers were assigned to candidate cluster members as listed
in Table \ref{obs-table}. They include objects identified via their X-ray 
emission (Garmire \& Garmire 2003; labeled G-number), completing the list 
with optical very low-mass stars and BD candidates (L\'{o}pez-Mart\'{\i} 
et al. 2005), and a large number of sky positions, required to 
perform a proper background subtraction in the inhomogeneous CrA nebula.
Since the selection was done independently of the Spitzer colors,
it is not biased toward objects with disks, and the sample contains
Class I, Class II, and Class III objects. The fibers have an aperture of
1.2 arcsec, so precise coordinates are necessary. All the X-ray detections 
were matched to 2MASS objects, or had their coordinates
transformed into the 2MASS system, to avoid fiber mismatching. For the
sources that were not detected with 2MASS nor Spitzer, there could be
some mismatching in the fiber position, since the X-ray coordinates
have in some cases uncertainties of the order of the fiber diameter
(see Section \ref{data-xray}), but this would in any case affect only
very few cases, if any. A total of 56 fibers were assigned to objects, 
and 70 fibers were positioned to measure the sky.

The spectra were reduced using IRAF\footnote{IRAF 
is distributed by the National Optical Astronomy Observatories,
which are operated by the Association of Universities for Research
in Astronomy, Inc., under cooperative agreement with the National
Science Foundation.} standard tasks within the packages
\textit{noao.imred.ccdred} and \textit{noao.imred.specred}.
A total of 3$\times$2700 s exposures were taken with the
gratings L682.2 and L773.4, and 2$\times$2700 s with the 
grating L881.7, in order to minimize the effects of cosmic ray hits.
The spectra were bias corrected, combined, and then extracted and
flat fielded using the specific task for multifiber spectroscopy,
\textit{dofibers}. This task was also used to determine the wavelength
calibration, using the available ThAr lamp. Due to the variations
of the sky over the whole region, as well as some stray light
affecting part of the fibers on the CCD, the sky subtraction was
done individually for each object. The different sky spectra were
divided in groups according to their brightness and the location 
on the CCD. Each group had a minimum of 20 individual sky spectra,
so the noise of the images was not substantially increased by
the sky subtraction. The sky emission lines were corrected
from each spectrum by subtracting the appropriate combined sky spectrum.
Due to the differences in transmission between fibers, and since it
is not required for the analysis we present here, no flux
calibration was performed. In order to match the spectra over the
entire wavelength range, we scaled the instrumental flux in the regions
where the spectra overlap ($\sim$80\,\AA\ between gratings L682.2 and
L773.4, and $\sim$140\,\AA\ between L773.4 and L881.7). 

Finally, the presence of emission and absorption lines (especially 
H$\alpha$ and the youth indicator Li I at 6708\,\AA) were measured
with IRAF task \textit{splot}. Of all the 56 objects observed with 
FLAMES, 17 were not detected at optical wavelengths, being most likely highly 
extincted stars or protostars (Class I/Class 0 objects), or (mainly the 
objects outside the high extinction cloud) extragalactic objects. Since none 
of the non-detected objects is detected in the 2MASS survey, their optical
magnitudes are likely to be very faint, and 
therefore, we believe they are rather non-detections than fiber mismatching.
One of the objects detected (G-115) is confirmed to be a quasar at z=0.62. 
For the rest, the H$\alpha$ line could be identified, and for 11 of them, 
the signal to noise ratio (S/N) was good enough to allow spectral typing 
(typically, S/N$>$100), or approximate spectral typing for 3 more objects 
(with S/N$\sim$30-100; see Section \ref{spectype}).

\subsection{IRS spectroscopy \label{data-irs}}

A total of 14 IR spectra were obtained with the low-resolution modules 
of the Infrared Spectrograph (IRS; Houck et al. 2004) on Spitzer, within 
our GO-3 program 30536. The observations took place between August 2006 
and October 2007. The objects were selected among the
X-ray members with IRAC and/or MIPS IR excesses, refining the
selection to ensure that the candidates were located in a zone
of moderate to low background. The observations were taken using
the high-precision optical pick-up, to ensure good centering of 
the object on the slit in a region with multiple IR sources and
strong emission from the ambient gas. The exposure times for
each object were calculated considering the IRAC and
MIPS fluxes (see Section \ref{data-archive}), and ranged between 
$\sim$200 s for the brightest sources to $\sim$105 min for the 
faintest ones (see Table \ref{obs-table}). In addition, we include the
spectrum of CrA-465, belonging to the GO3 program number 30540.

The spectra were reduced using the reduction schemes developed for the 
analysis of the data from the legacy programs ``Formation and Evolution of 
Planetary Systems" (FEPS) and ``Cores to Disks'' (c2d; Bouwman et al. 2006, 
2008 and references therein). This reduction package consist of automated 
routines, based on SMART (Higdon et al. 2004) and other IDL programs, 
to extract calibrated IRS spectra from the Spitzer pipeline basic 
calibrated data (BCD). We use a custom-size extraction window, fitted to the 
point spread function, which ensures better sensitivity than the default 
extraction aperture provided by the pipeline. For sources with high
S/N, the position of the aperture was fitted in the reduction process, in
order to minimize flux loses and ensure a good order matching.
In case of high-background regions, we cancel the sky in an efficient 
way by the subtraction of the opposite 
nod beams and extraction and subtraction of the two resulting sources. 
Bad pixel masks were created by examining the images, and included in the 
reduction. These data reduction techniques have produced 
high-quality IRS spectra for the FEPS program and for the $\eta$ Cha
very low-mass stars IRS program (Bouwman et al. 2006, 2008).

In the cases where the background levels were higher
and/or the objects were very faint (G-65, CrA-205, CrA-466, CrA-4110, 
CrA-4111) the background was examined at different locations
in the two nods (on both sides of the star and in the off positions),
selecting the best fitting one for the subtraction. In the case of
CrA-4110 and G-65, the presence of an extremely bright source and strong 
diffuse emission at the LL (long-low, 14-25\,$\mu$m) wavelengths made it impossible
to extract the source, so only the SL (short-low, 5-14\,$\mu$m) range is presented.

\subsection{IRAC, and MIPS complementary data\label{data-archive}}

To complete the disk information, especially for the objects without
IRS spectra, we searched the 2MASS database as well as the Spitzer 
Infrared Array Camera (IRAC; Fazio et al. 2004) and the Multiband
Imaging Photometer (MIPS; Rieke et al. 2004) data already being public in 
the Spitzer archive for the Coronet region. The complementary 2MASS
and Spitzer data for all sources observed with FLAMES and/or IRS are presented
in Table \ref{archive-table}. Since we did not find a list of IRAC and MIPS 
sources in the Coronet within the literature, the IRAC and MIPS data were 
downloaded using \textit{Leopard}. The data covering our objects were obtained from 
programs 6 (AORs 3650816 and 3664640), 30784 (AOR 17672960), and 248 (AOR 13469696). 
These programs include two IRAC fields ($\sim$0.3$\times$0.3 deg) with small
overlap in the standard map mode (12s exposure for AOR 3650816, and 6$\times$12s
for 17672960), a small MIPS 24\,$\mu$m map ($\sim$7$\times$7 arcmin, 3$\times$10s 
exposures), and large MIPS 24 and 70\,$\mu$m scan maps ($\sim$0.6$\times$1 deg, 
exposure time 3.67s). The very large areas covered by IRAC and MIPS ensure
that all the X-ray and optical candidates are included in at least two
IRAC channels (95\% of the objects are covered by IRAC maps in all the four
channels) and in the large MIPS map.

We constructed the mosaics using the software MOPEX, starting on the BCD data, and 
using the standard parameters in the Spitzer Mosaicker Manual 
\footnote{http://ssc.spitzer.caltech.edu/postbcd/onlinedocs/intex.html}. A 10\% accuracy 
in the photometry is sufficient for the aim of this work, and the main sources of error 
in the region are the variable background and the photon counts for the faintest objects, 
so we did not correct for array location dependence. 

We obtained aperture photometry using the IRAF task \textit{phot} in the package 
\textit{noao.digiphot.apphot}. For the four IRAC channels, we used a 5 pixel aperture 
with a background annulus 5-10 pixels, with the corresponding aperture corrections 
listed in the IRAC Data Handbook (1.061, 1.064, 1.067, and 1.089 for the
3.6, 4.5, 5.8 and 8.0\,$\mu$m channels, respectively). For the MIPS 24\,$\mu$m 
photometry, we used a 5 pixel aperture with background annulus 15-19 pixels
and an aperture correction of 1.167 (MIPS Data 
Handbook\footnote{http://ssc.spitzer.caltech.edu/mips/dh}). These values
had been successfully applied to our Spitzer IRAC and MIPS photometry of Tr 37 and the 
Tr 37 globule (Sicilia-Aguilar et al. 2006a). For the few sources detected 
at 70\,$\mu$m, we used a 7.5 pixel aperture, background measured from 10-15 pixels, and 
an aperture correction 1.295, following the MIPS Data Handbook. In 
case a source was detected with equally good quality in two programs, we use 
the average magnitude (note that differences between two measurements are
always small and fully consistent with the photometric errors in 
Table \ref{archive-table}). The main limits of the IRAC and MIPS photometry are
the contamination of large parts of the maps by the saturated emission of 
nearby bright sources (in particular, R CrA) and extended nebular emission.
For objects with moderate extinction (A$_V <$15 mag), the IRAC data is complete
down to photospheres of M7 type objects, and down to M4-M5  photospheres
at 24\,$\mu$m. The detection limits for MIPS are $\sim$1-2 mJy (magnitude $\sim$9.5-10) 
at 24\,$\mu$m, and $\sim$80 mJy (magnitude $\sim$2.5) at 70\,$\mu$m. The IRAC and MIPS 
data are also in very good agreement with the fluxes measured in the IRS spectra.

\section{Analysis \label{analysis}}

\subsection{Spectral types and extinction \label {spectype}}

\begin{figure}
\plotone{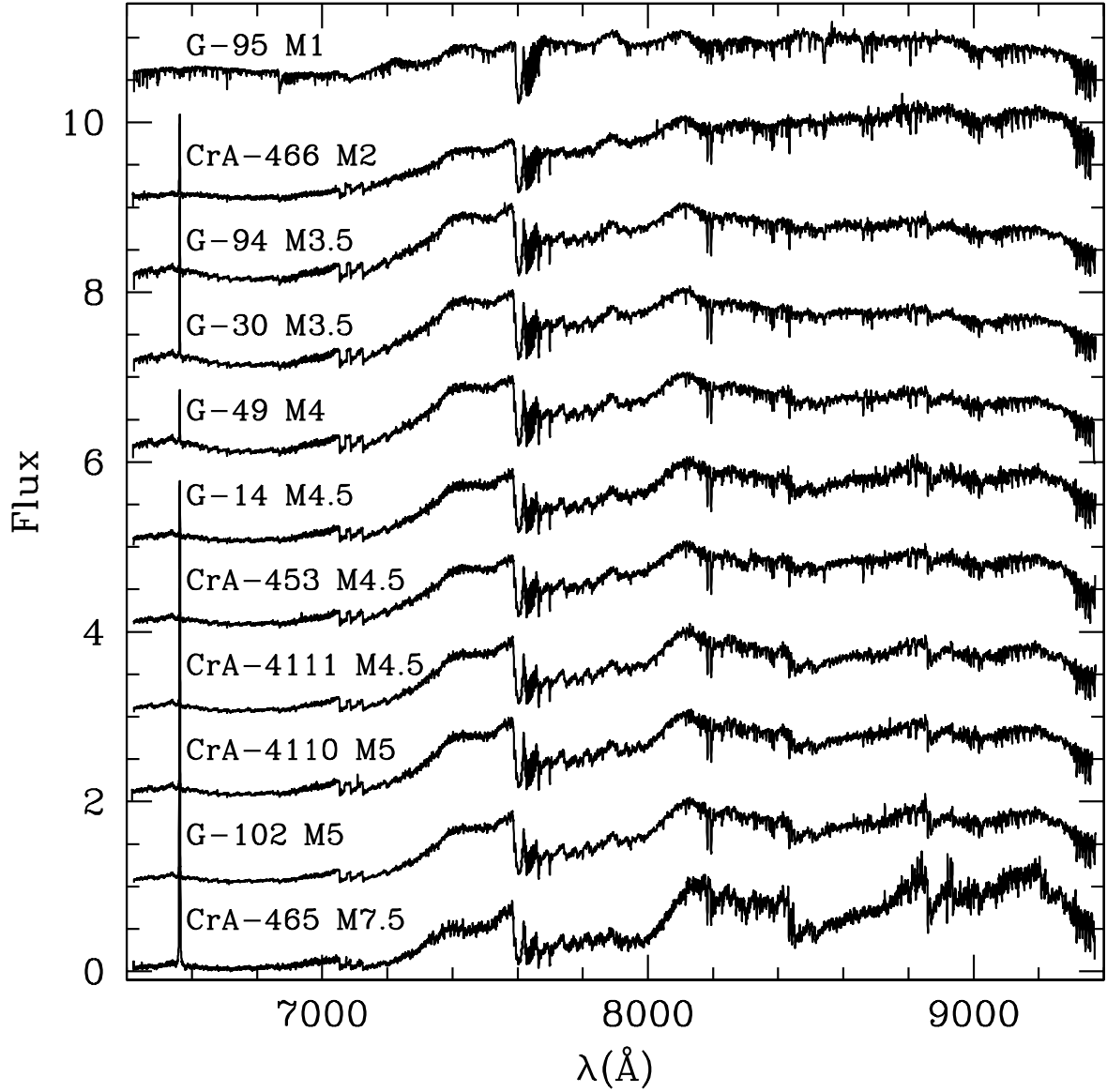}
\caption{FLAMES spectra of some of the Coronet cluster objects. The spectra
have been scaled and shifted in order to display them together. \label{spectra-fig}}
\end{figure}

The first step on the analysis of the FLAMES spectra was to derive spectral types 
and extinction values for the observed objects. Given that all the spectra showed 
the typical TiO and VO bands of M-type stars (see Figure \ref{spectra-fig}), we 
used a combination of spectral typing indices for M stars, containing four PC 
(pseudo-continuum) indices from Mart\'{\i}n et al. (1996), namely PC1, PC2, PC3 
and PC4, and the R1, R2, R3, TiO 8465 indices in Riddick et al. (2007), 
listed in Table \ref{index-table}. In case the spectral type was M5 
or later, the results were refined by using the indices VO 2 and VO 7445 
(Riddick et al. 2007). We avoided the indices that can be affected by atmospheric 
absorption (O$_2$ and H$_2$O bands), and those falling near the extremes of the 
FLAMES spectra. All the indices were obtained by measuring the ratio of average 
fluxes in two or more places in the spectra (Table \ref{index-table}).
We use the calibrations derived from collections of standard stars and BD
listed in Mart\'{\i}n et al. (1996) and Riddick et al. (2007), except for 
the PC1 and PC2 indices, for which we fitted a linear relation, including 
all the objects listed in Mart\'{\i}n et al. (1996), in order to improve the
quality of the fit (see Table \ref{index-table}). 

The PC indices have the 
disadvantage that they compare zones of the spectrum separated by $\sim$500-1500 \,\AA,
which makes them sensitive to extinction. The Riddick et al. indices compare 
fluxes which are not more than 200 \,\AA\ apart, so the effect of extinction is 
negligible. Gravity can lead to variations as well. The calibration in 
Riddick et al. (2007) is constructed for pre-main sequence objects 
with ages similar to the members of the Coronet cluster, so the errors due 
to differences in gravity are expected to be minimal. We did not find significant 
differences between the spectral types obtained from extinction- or gravity-sensitive
indices and others, so we believe the results are robust. Combining
several indices for each object, the typical errors in the spectral type 
are around 0.5-1 subtypes. For stars M2 or earlier, for which the described indices 
are not very precise, we made a visual comparison of the spectra with those of standard
stars in order to refine the classification. For the objects with poor S/N ($\sim$30-100), 
for which the indices were not accurate, we obtained an approximate spectral type by visual 
comparison to standard spectra, checking the 2MASS colors as well to reduce the 
uncertainties. Our results are in good agreement with the spectral types or approximate 
spectral types listed by L\'{o}pez-Mart\'{\i} et al. (2005) and Forbrich \& Preibisch (2007).

\begin{figure}
\plotone{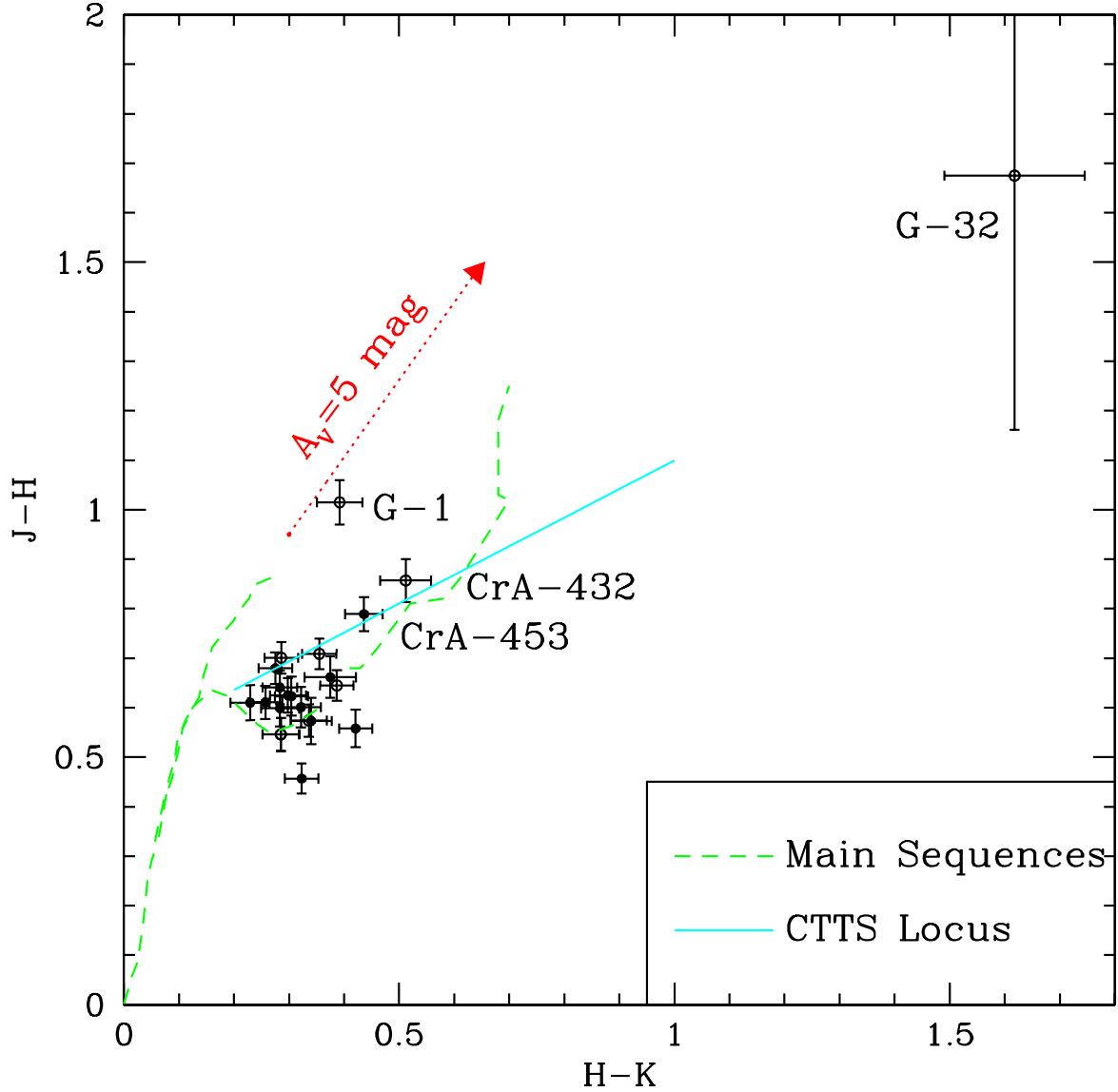}
\caption{J-H vs. H-K diagram for the very low-mass objects in the Coronet
Cluster. The main sequences of dwarfs, giants, and BD are displayed as a dashed line
for comparison (Bessell et al. 1998; Kirkpatrick et al. 1996), together with the CTTS
locus stars (Meyer et al.1997, solid line), and a reddening vector for
A$_V$=5 mag. The objects with known extinction has been corrected and are represented by
filled symbols. The objects for which the extinction is not known are not corrected and
are represented as open circles. All the objects with small photometric errors
have colors consistent with M-type stars or BD with no or little near-IR excess. 
The objects that stand out of the M0-M6 track (either because of their extinction 
or because of having later spectral types) have been labeled. \label{jhk-fig}}
\end{figure}

For the objects with known spectral types, the extinction was estimated
using the J-H and J-K colors observed with 2MASS, and comparing to the
standard colors for stars (Bessell et al. 1998) and 
for BD (Kirkpatrick et al. 1995). For each object, the
extinction A$_V$ was derived from E(J-H) and E(H-K).
Following Bessell \& Brett (1988), and assuming a typical galactic extinction
law, A$_V$=10.87 E(J-H) and A$_V$=5.95 E(J-K) mag. The final extinction was
obtained as the average, with errors given as the standard deviation. 
The results are similar (within 5\%) to those using the relations 
in Cardelli et al. (1989). Since many of the
studied objects have circumstellar material, part of the matter causing the
extinction could be processed to some degree, which may result in non-standard
extinction laws. Nevertheless, the agreement between the extinction obtained
from both J-H and J-K suggest that the deviations, if any, are small. 
The location of the objects within the J-H vs. H-K diagram 
(Figure \ref{jhk-fig}) is in good agreement with the spectral types (M stars 
and a BD, CrA-465) as well as with the standard extinction law. The spectral
types and extinctions are listed in Table \ref{lines-table}.

For the objects with no optical spectra, approximate spectral
types could be inferred only from the SED fitting. For those two classes, 
we can estimate an approximate extinction using the JHK diagram 
(Figure \ref{jhk-fig}), since their disks produce no significant excess
at those wavelengths. From this, we can conclude that G-1 is a late-K or
early-M star with A$_V \sim$3.5 mag (Forbrich \& Preibisch 2007 classify it
as a potential M-type star), and G-32 is a deeply embedded Class II object 
whose uncertain 2MASS photometry does not allow any extinction
estimate. All the rest of objects are consistent with M stars and maybe a
massive BD (M6-M8, CrA-432), with extinction A$_V <$2 mag. 

\subsection{Dust mineralogy \label{disks}}

\begin{figure}
\plotone{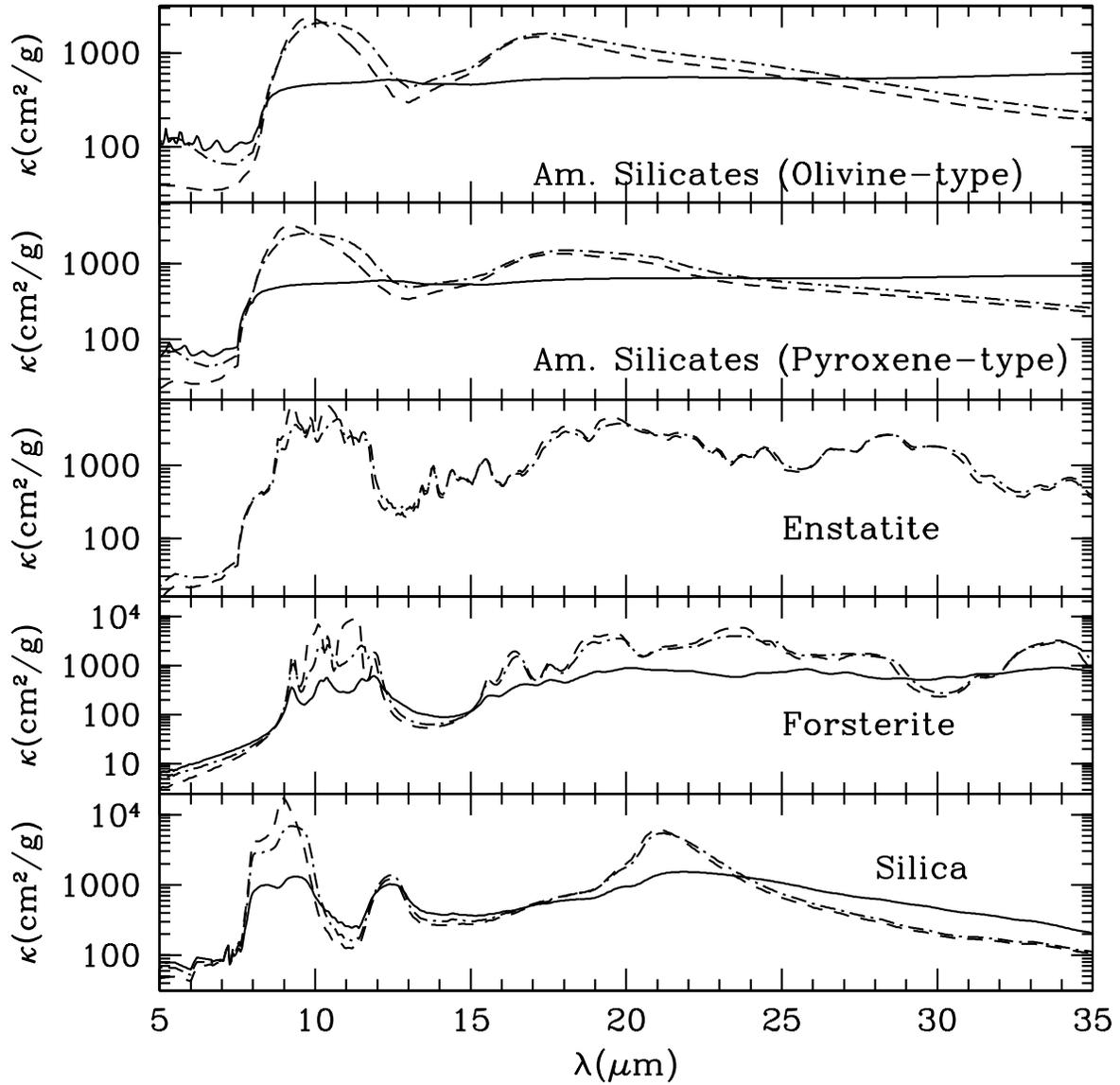}
\caption{Mass absorption coefficients ($\kappa$) for the materials used in the 
five-particle model silicate fitting routine (Dorschner et al. 1995; Servoin \& Piriou 1973; 
J\"{a}ger et al. 1998; Henning \& Mutschke 1997; Preibisch et al. 1993).
Two amorphous silicates (with olivine and forsterite stoichiometry), plus
enstatite, forsterite, and silica, are used. For each material, we include three
grain sizes: 0.1\,$\mu$m (dashed line), 1.5\,$\mu$m (dotted-dashed line), and 6.0\,$\mu$m
(solid line), except for enstatite, for which we have only 0.1\,$\mu$m (dashed line) and
1.5\,$\mu$m (dotted-dashed line) sized-grains. \label{qval-fig}}
\end{figure}

Some of the disks observed show silicate emission features in the 8-13$\mu$ and 
20-30\,$\mu$m range in their IRS spectra. These features are produced by the small grains 
($\leq$6\,$\mu$m) in the warm ($\sim$150-450 K), optically thin, disk atmosphere. 
In order to analyze the dust mineralogy in this layer, we 
fitted the spectra  using the Two-Layer Temperature Distribution (TLTD)
spectral decomposition routines developed by Juh\'{a}sz et al. (2008). This model
is an improvement of the single- and two-temperature fitting methods
(Bouwman et al. 2001; van Boekel et al. 2005). It reproduces the silicate 
emission using the sum of a multicomponent continuum  (star, inner rim, 
disk midplane) assuming that the region where the observed radiation originates 
(both optically thin and thick) has a distribution of temperatures instead of a single
one. The routine includes the dust species for which we
have evidence of optically thin emission in the IRS wavelength range.
These are five different dust species (amorphous silicates with olivine and pyroxene 
stoichiometry, forsterite, enstatite, silica) with sizes 0.1, 1.5, and 6.0\,$\mu$m 
(see Figure \ref{qval-fig}; Dorschner et al. 1995; Servoin \& Piriou 1973; 
J\"{a}ger et al. 1998; Henning \& Mutschke 1997; Preibisch et al. 1993), except for 
the enstatite grains, for which we only include 0.1 and
1.5\,$\mu$m grains. This is due to the fact that there is no evidence of large 
enstatite grains in disks (Bouwman et al. 2001; Sicilia-Aguilar et al. 2007), and
also because the inclusion of a large enstatite component adds a spurious source
of continuum that masks the more abundant amorphous grains, and extra features
in the 13-14\,$\mu$m region that are rather produced by gaseous components (C$_2$H$_2$
and HCN; Carr \& Najita 2008; the spectra of CrA-466, CrA-432, and G-87
show marginal detections of gas lines in this region). Amorphous carbon grains 
with sizes 0.1, 1.5 and 6.0\,$\mu$m are also included as a featureless contribution
to the continuum. The mass absorption coefficients are derived from
the material optical constants using the theory of distribution of hollow
spheres for the crystalline dust (Min et al. 2005\footnote{Previous investigations
show that this scattering model provides a relatively good fit to the silicate
features (e.g., Bouwman et al. 2001). The application of more complicated dust
models (e.g. Henning \& Stognienko 1996; Voshchinnikov \& Henning 2008) is beyond
the scope of this paper.}), and the classical Mie 
theory for spherical particles for the amorphous dust. 

The observed flux-density at a given frequency is given by 

\begin{eqnarray}
F_\nu = F_{\nu, {\rm cont}} & + & \sum_{i=1}^N\sum_{j=1}^MD_{i,j}\kappa_{i,j}
\int_{\rm{T_{\rm a, max}}}^{\rm{T_{\rm a, min}}}\frac{2\pi}{d^2}B_\nu(T){T}^{\frac{2-qa}{qa}}dT
\label{eq:1}
\end{eqnarray}

where, $N$ and $M$ are the number of dust species and of grain sizes, respectively.
The mass absorption coefficient of the dust species $i$ and grain size $j$ 
is represented by $\kappa_{i,j}$. $B_\nu(T)$ is the Planck-function, $qa$ is the power 
exponent of the temperature distribution and $d$ is the distance to the source. The 
subscript $a$ in the integration boundaries refers to the disk atmosphere. The continuum 
emission ($F_{\nu, {\rm cont}}$) is given by

\begin{eqnarray}
F_{\nu, {\rm cont}} = D_0 \frac{\pi R_\star^2}{d^2} B_\nu(T_\star)&+& D1\int_{\rm{T_{\rm r,max}}}^{\rm{T_{\rm r, min}}}\frac{2\pi}{d^2}B_\nu(T){T}^{\frac{2-qr}{qr}}dT \\
&+& D2\int_{\rm{T_{\rm m,max}}}^{\rm{T_{\rm m, min}}}\frac{2\pi}{d^2}B_\nu(T){T}^{\frac{2-qm}{qm}}dT.
\label{eq:2}
\end{eqnarray}

The first term on the right hand side describes the emission of the star the while the  
second and third terms describe the radiation of the inner rim and the disk midplane respectively.  
The meanings of the different parameters are summarized in Table \ref{silparam-table}. 
The parameters include the distance d (150 pc; Marraco \& Rydgren 1981; Knude \& Hog 1998) and
the radius and effective temperature of the star $R_*$ and $T_*$  (estimated from the spectral 
type and luminosity of the sources). The highest temperature of each component (disk atmosphere,
inner rim, midplane) is fitted, and the lowest temperature is calculated assuming that the
contribution of the annulus with the corresponding temperature to the total flux is
higher than 0.1\%. Small variations in the disk model parameters 
(inner and outer disk temperatures, temperature exponents) do not produce significant
changes in the extracted mineralogy, as the bulk of the silicate emission comes from 
a region with T$\sim$150-450 K, and no dust exists beyond the dust sublimation temperature 
($\sim$1200 K). The exponents of the temperatures (qa, qm, qr), and the coefficients 
of each contribution ($D_0$, $D_1$, $D_2$, $D_{i,j}$) are fitted (see details 
in Juh\'{a}sz et al. 2008 and more information in Table \ref{silparam-table}). 
To account for the errors in the IRS spectra, the fit of each object was repeated 
100 times, adding random Gaussian noise to the original spectrum. The final 
silicate composition is obtained as the average of the whole set, and the errors 
are derived from the standard deviation, considering separately the positive and
negative directions (Van Boekel et al. 2005; Sicilia-Aguilar et al. 2007). 
The estimated errors include therefore only the 
component due to the S/N and measurement errors in the spectra. Other sources of error,
like those related to the presence of materials that are not included in our
set of optical constants, or different grain shapes or chemical composition
of the grains, could introduce systematic deviations that are harder to quantify and 
will be studied in detail in the future (Juh\'{a}sz et al. in prep.).

\begin{figure}
\plotone{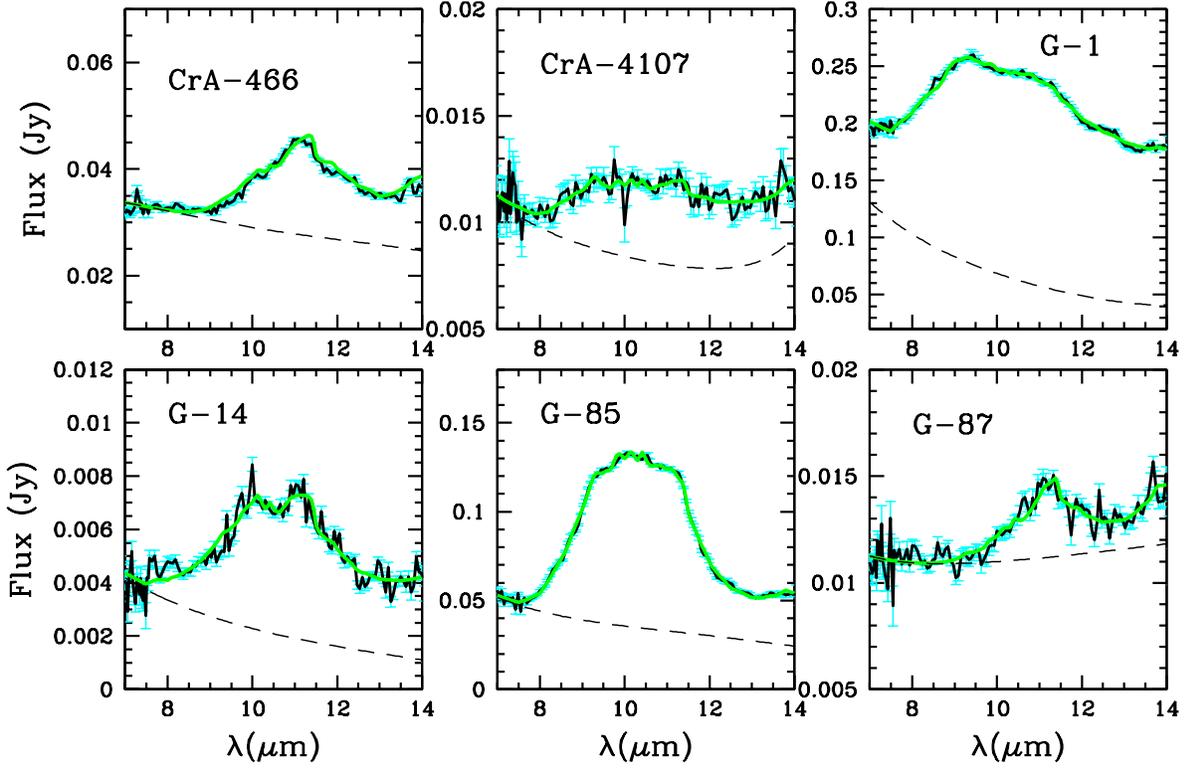}
\caption{Silicate emission and fit of the objects with excess emission in the
10\,$\mu$m region. The spectra with noise in the fitted region (7-14\,$\mu$m) are 
displayed. The final fit is represented by a thick line, and the continuum level
is marked as a dashed thin line.\label{silfit-fig}}
\end{figure}

\begin{figure}
\plotone{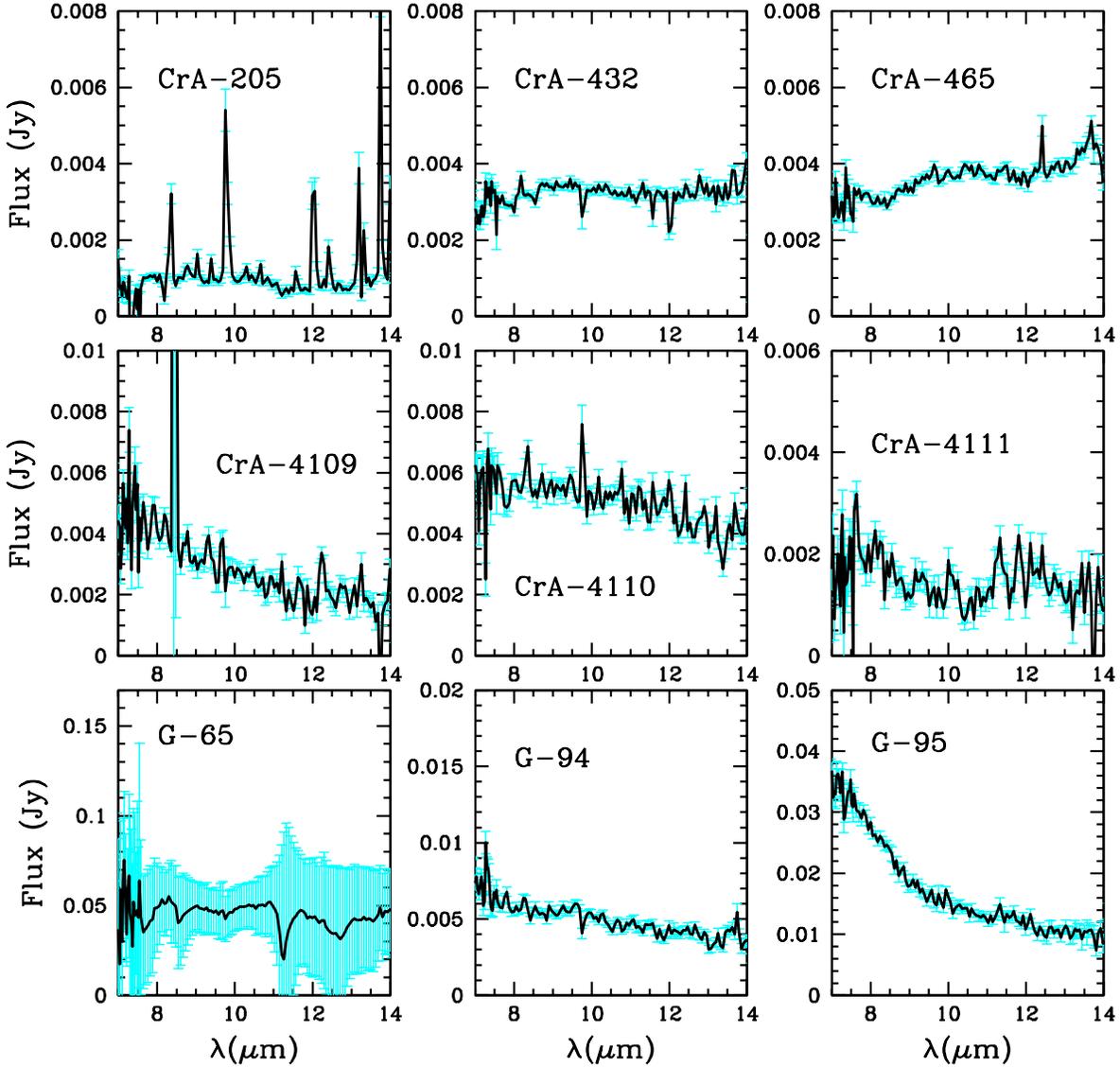}
\caption{10\,$\mu$m region in the objects with no evident silicate emission.
Note that CrA-432 and Cra-465 do not show any silicate emission,
but just excess emission starting at $\sim$10\,$\mu$m. CrA-205 shows multiple
lines that are gas lines from the nebula and bad pixels, and cannot be totally 
corrected due to the very low emission from the source.
The large errors in G-65 are due to the very bright background near R CrA.
\label{nosil-fig}}
\end{figure}

\begin{figure}
\plotone{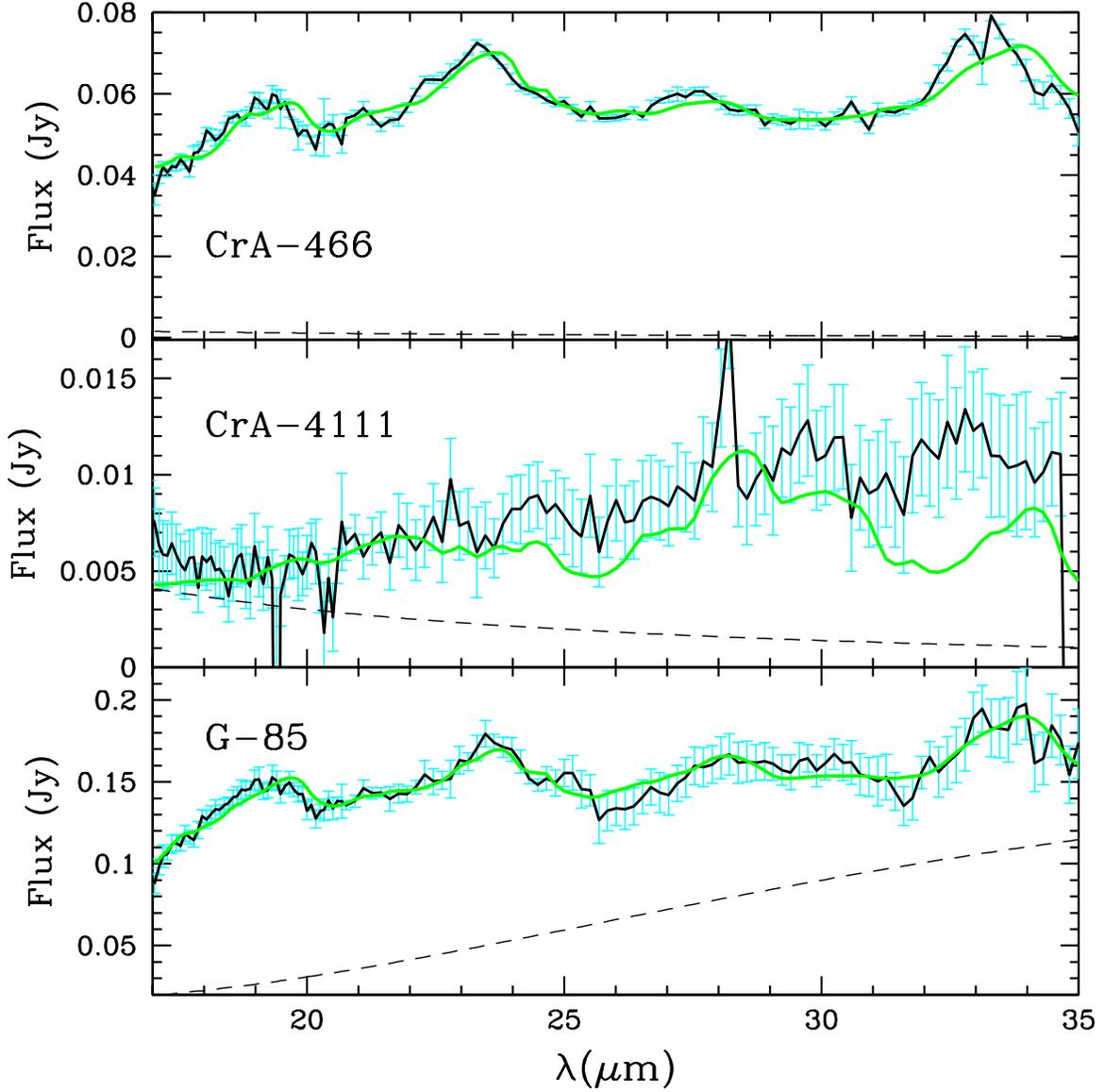}
\caption{Silicate emission and fit in the 17-35\,$\mu$m range. 
The small offsets between the features and the model in CrA-466 may
suggest a difference in composition. The spectra with noise in the fitted region are 
displayed. The final fit is represented by a thick line, and the continuum level
is marked as a dashed thin line.
\label{longsilfit-fig}}
\end{figure}

The silicate feature at 8-13\,$\mu$m is detected in 6 objects out of 13 
(CrA-466, CrA-4107, G-1, G-14, G-85, and G-87;  Figure \ref{silfit-fig}).
A zoom on the 10\,$\mu$m region for the objects without evidence of silicate 
feature is shown in Figure \ref{nosil-fig}. 
The S/N is in all cases good enough to determine the presence or absence of 
silicate emission. The S/N is typically worse at longer wavelengths, so 
the silicate features in the  20-30\,$\mu$m range could be identified only
in three of the spectra (CrA-466, CrA-4111, G-85; Figure \ref{longsilfit-fig}), 
even though one of them (CrA-4111) shows no silicate emission in the 10\,$\mu$m 
region. Since the silicate signatures at 8-13\,$\mu$m and 20-30\,$\mu$m are dominated 
by material at different temperatures and thus, different radial distances from the
star, having probably different compositions and grain sizes, it is not 
recommended to fit the whole spectrum with a single model. Therefore, we fitted the 
7-14\,$\mu$m wavelength range for the shorter wavelengths,
which offers the best compromise taking into account the optical constants and radial
changes (Juh\'{a}sz et al. 2008) and the quality of the spectra. Juh\'{a}sz et al. (2008)
suggest to use the 7-17\,$\mu$m region for fitting Herbig AeBe stars, but most of the spectra 
here have low S/N, so small errors in the order matching at $\sim$14\,$\mu$m produce small
offsets in the final spectra and/or increased noise, which affects the final fit. Although
the routine used for fitting includes a re-centering of the aperture that substantially
improves order matching, the result is not perfect, specially for the weaker objects.
For the longer wavelength features, we fit the 17-35\,$\mu$m region. The results of the fitting
procedure are listed in Table \ref{sil-table}, and displayed in Figures \ref{silfit-fig} 
(7-14\,$\mu$m region) and \ref{longsilfit-fig} (17-35\,$\mu$m region). The disks show 
typically a mixture of crystalline and amorphous silicates with different grain sizes, 
and a large variety of abundances, with no evident correlation between disk/stellar properties
and silicate emission (Table \ref{sil-table}). Evidence of potential
differences in chemical composition of the grains (for instance, the presence of a fraction 
of Fe silicates instead of Mg silicates) could cause systematic offsets of the lines 
similar to those seen in CrA-466 at long wavelengths (Figure \ref{longsilfit-fig}).
Differences in grain porosity could also lead to similar effects (Voshchinnikov \&
Henning 2008). 

Among the spectra with silicate emission in the 7-14\,$\mu$m region and low errors in the 
fitted grain composition (CrA-466, G-1, G-14, G-85), we observe a large variety of 
crystalline fraction, going from $\sim$3-5\% for G-14 and G-1, to $\sim$20-30\% for CrA-466
and G-87 (Table \ref{sizecryst-table}). The crystallinity fraction of CrA-466 and G-87 are 
higher than observed for the solar-type stars in the 4 Myr-old cluster Tr 37 
(Sicilia-Aguilar et al. 2007), suggesting strong grain processing. High crystallinity 
fractions compared to those measured for disks around solar-type stars are typical in 
the disks around low-mass stars and BD (Apai et al. 2005). Very large crystalline grains 
are rare, probably because they form aggregates together with amorphous grains 
(Bouwman et al. 2006), but large amorphous grains ($\sim$6\,$\mu$m) are found in many cases 
(e.g., Sicilia-Aguilar et al. 2007). Therefore, strong grain growth may result in part 
of the amorphous dust being too large to produce detectable features in the IRS spectra,
while the crystalline grains may still be visible and seem to be more abundant.
Inclination effects can also lead to variations in the silicate feature for
objects close to edge-on, but given that these disks are very flattened, strong
variations are expected to be rare.

In the long-wavelength region, only the spectra CrA-466 and G-85 could be fitted
with an uncertainty $<$10\% (Figure \ref{longsilfit-fig}). 
CrA-466 is characterized by mostly large amorphous grains with 
pyroxene stoichiometry, and a small fraction of small forsterite and large silica grains.
The crystalline fraction is therefore smaller ($\sim$14\%) than in the 10\,$\mu$m region,
which could suggest the presence of larger quantities of small amorphous grains at large
radii. For G-85, there are no significant differences between the 7-14\,$\mu$m
and the 17-35\,$\mu$m region in crystallinity fraction, although the grain sizes tend to
be smaller in the longer wavelength range, and the amorphous silicates are dominated
by pyroxene-type components at shorter wavelengths, but olivine-type grains
in the long wavelength range. The presence of strongly processed crystalline grains at long 
wavelengths suggests either some radial mixing at some stage in the evolution, or the importance
of shocks in thermal annealing. Nevertheless, observations of larger samples of objects 
are required to draw statistically significant conclusions. 

As we mentioned, half of the disks have no warm (150-450 K) small ($<$6\,$\mu$m) 
grains in the disk atmosphere. For the transition objects (TO) or disks with inner 
holes, this may be in fact the effect of lack of dust in a physical clean hole 
(0.1-1 AU, depending on the spectral type). 
Nevertheless, 3 of the TO do show silicate emission (CrA-466, G-14, and G-87), 
and 3 of the disks without near-IR inner holes lack the feature (CrA-432, 
CrA-465, CrA-4110), so we probably require other physical conditions (strong grain
growth, flattened disks) to explain this behavior. The lack of silicate features in
flattened (but non-transitional) disks has been also observed in the 4 Myr-old
cluster Tr 37 (Sicilia-Aguilar et al. 2007). The S/N is typically lower 
($\sim$10-20) in the 20-30\,$\mu$m range, and considering that the features are
much weaker in this part of the spectrum, we cannot confirm nor rule out the
presence of small silicate grains in the colder parts of the disk for the rest of
the objects.

\section{Discussion \label{discussion}}

\subsection{Disks around very low-mass objects: ``Transition'' disks?\label{lowmass}}

\begin{figure}
\plotone{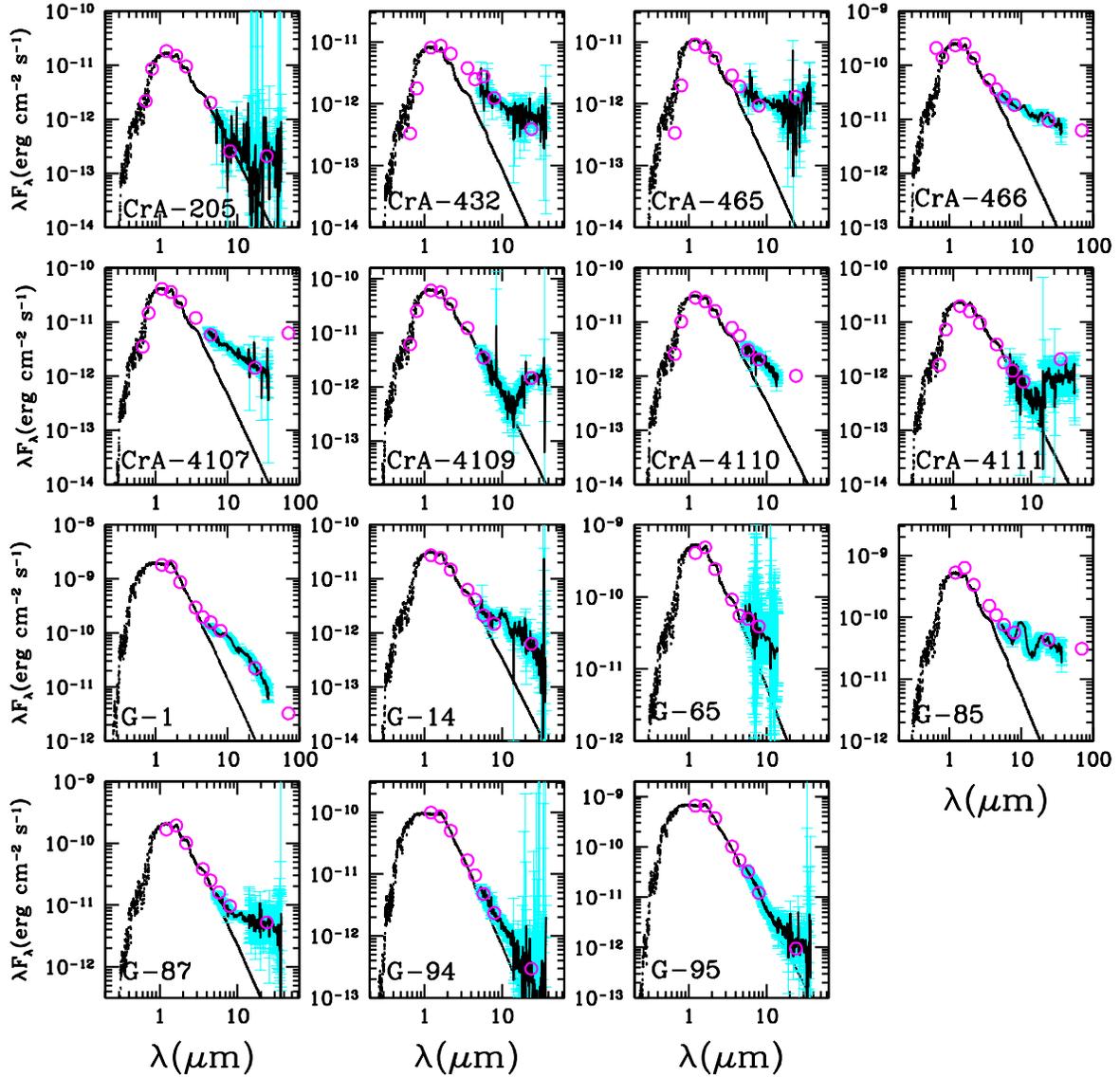}
\caption{SEDs for the objects observed with IRS. A stellar photosphere of a similar spectral type
is displayed for comparison. All the fluxes with known extinction have been dereddened,
including G-1 (A$_V$ $\sim$ 3.5 mag derived from the JHK diagram). The data points 
include optical R and I photometry (L\'{o}pez-Mart\'{\i} et al. 2005),
2MASS, IRAC, and MIPS observations. \label{irssed-fig}}
\end{figure}

\begin{figure}
\plotone{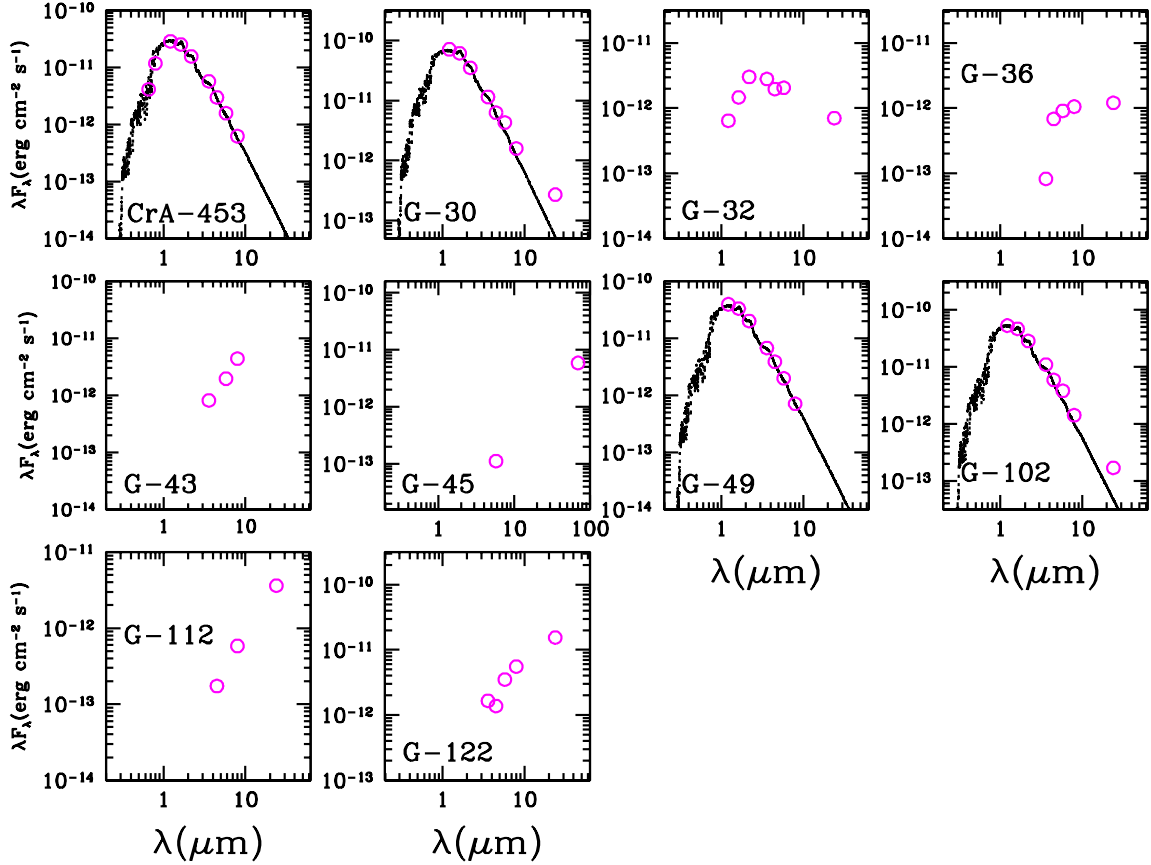}
\caption{SEDs for the objects not observed with IRS. In the case of Class III and
Class II objects with low extinction, a stellar photosphere of a similar spectral type
has been displayed for comparison.  The data points include optical R and I photometry 
(L\'{o}pez-Mart\'{\i} et al. 2005), 2MASS, IRAC, and MIPS observations.
\label{sed-fig}}
\end{figure}

Since this study targeted candidate young objects based
on X-ray or optical photometry, which are not biased towards systems with 
IR excesses from circumstellar disks, it is possible to estimate the disk 
fraction for low- and very-low mass objects in the cluster. As mentioned in
Section \ref{data-xray}, the X-ray survey is complete down to the spectral range
considered here ($\sim$M8) within the Chandra field, which is totally covered by
the IRAC and MIPS maps (Section \ref{data-archive}). The stars/BD
with SEDs consistent with young Class II and Class III sources and 
TO, clearly identified via JHK and IRAC/MIPS photometry
(Figures \ref{irssed-fig} and \ref{sed-fig}) sum up to 20 objects. They 
are all  M0 stars or later, except maybe G-1, which could be a late-K or 
early-M star. Therefore, this sample contains objects with later spectral 
types than the Haisch et al. (2001), Sicilia-Aguilar et al. (2006a), 
and Hern\'{a}ndez et al. (2007a) studies, 
which were done for solar-type (G, K, early M) stars. 

\begin{figure}
\plotone{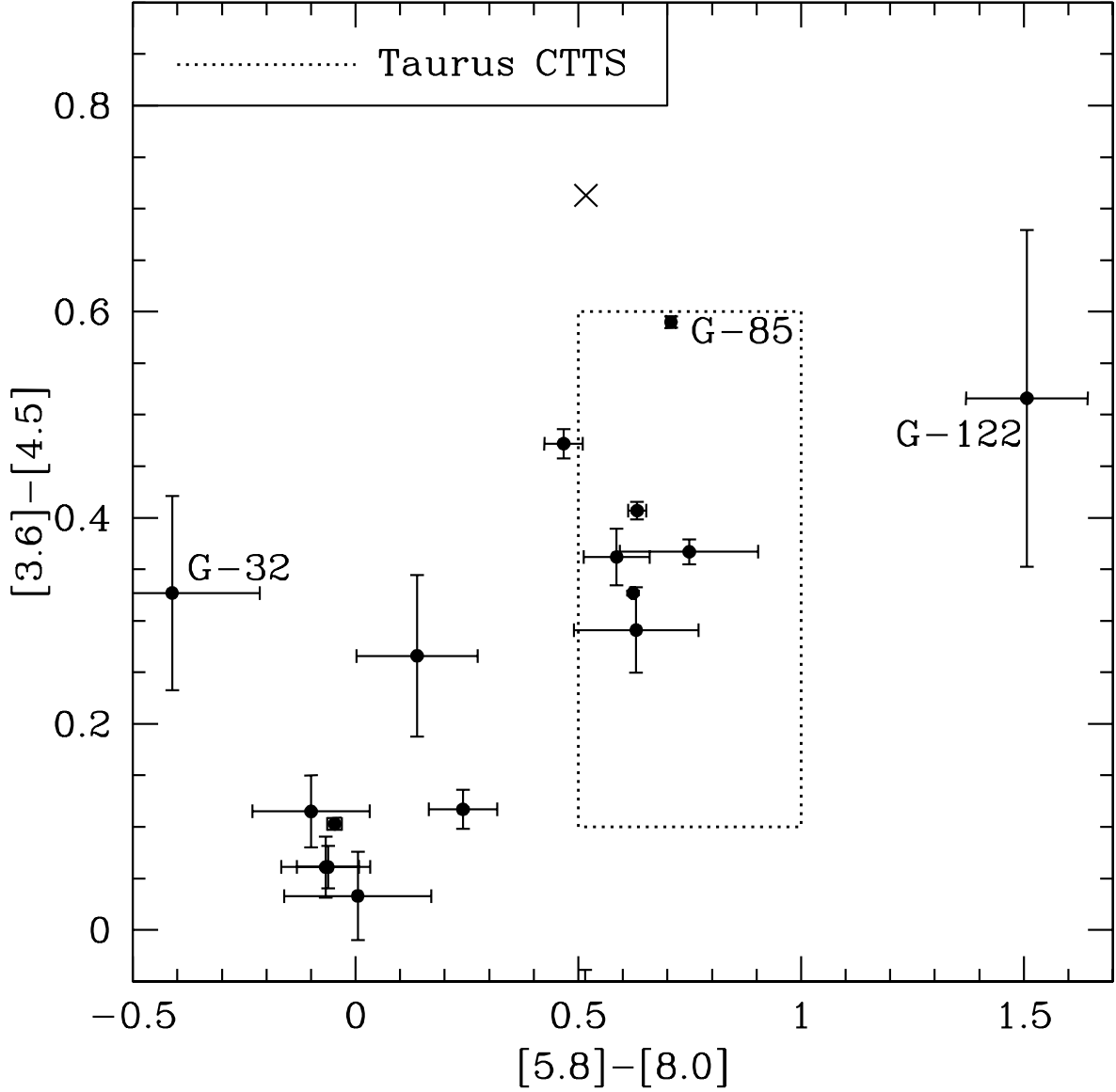}
\caption{IRAC color-color diagram for the very low-mass objects in the Coronet cluster.
CTTS-like objects with disks tend to appear within the box, which shows the
location of typical Taurus CTTS (Hartmann et al. 2005). WTTS-like objects
reside near the 0,0 point. Disks around these very low-mass objects tend to be flatter
than their higher-mass counterparts in Taurus. The X marks the quasar G-115, which
falls in the region for normal Class I sources or very flared Class II disks. \label{irac-fig}}
\end{figure}

The IRAC color-color diagram (Figure \ref{irac-fig}) shows colors consistent with 
typical Class II disks in Taurus, together with some transitional disks and 
Class III-like objects. From the shape of the SED and the wavelength at which the IR 
excess starts, 7 are Class II objects (with IR excess starting before 6\,$\mu$m, 
CrA-432, CrA-465, CrA-4107, CrA-4110, G-1, G-32, and G-85), 6 are Class III objects (no 
IR excess, or only marginal 24\,$\mu$m excess detected, CrA-453, G-30, G-49, G-94, 
G-95, and G-102), and 7 are TO (with no IR excess at wavelengths shorter than 
6\,$\mu$m, and sometimes, longer; CrA-205, CrA-466, CrA-4109, CrA-4111, G-14, G-65, 
G-87). This results in a disk fraction $\sim$70\%, of which 50\% of 
the disks are TO. Note that most of the Class II objects (all but CrA-432 and G-32) 
have no excess at JHK nor at 3.6\,$\mu$m, so their disks are very different from 
the typical Class II disks with JHK colors falling within the ``CTTS locus"
(Meyer et al. 1997). The disk fraction for the Coronet cluster is slightly lower 
than expected  for solar-type stars in a $\sim$1 Myr-old region with significant cloud 
material and several embedded Class I and Class 0 objects (usually $\sim$80\% in JHKL 
surveys; Haisch et al. 2001; Hern\'{a}ndez et al. 2007a). 

There is a controversy about the timescales for disk dissipation around very low-mass 
stars and BD. Most studies suggest a longer lifetime or larger disk fractions for 
the very low-mass objects compared to solar-type stars (Sterzik et al. 2004; 
Bouy et al. 2007), but some other studies have claimed 
similar timescales (Luhman et al. 2005; Damjanov et al. 2007) and even 
shorter lifetimes for the M-type stars (Lada et al. 2006). An important problem 
estimating disk fractions is the wavelength at which the excess is measured,
and the completeness of the observations at the given wavelength. In the Coronet 
Cluster, only $\sim$1/3 of the disks show excess at JHKL wavelengths similar to 
the excesses tracked by Haisch et al. (2001). This is a significantly lower fraction 
than for solar-type stars with the same age, for which only 5-10\% of the disks 
lack near-IR excesses (e.g., Taurus; Hartmann et al. 2005), and has been
pointed out already for BD (Luhman et al. 2007). Lada et al. (2006) 
calculated their final disk fractions for ``optically thick'' disks only, which 
excludes all the TO and the flattened (``anemic'') disks. The fact that a large 
fraction of the disks around the very low-mass Coronet cluster members are transitional 
or ``anemic'' disks, using the definitions in Lada et al. (2006), can explain the 
controversies in disk fraction for very low-mass objects. In addition, the 
morphology of the SEDs in the Coronet requires observations at wavelengths longer 
than $\sim$3.6\,$\mu$m to detect the presence of circumstellar material accurately. 
The lack of complete long-wavelength IR data can also contribute to the uncertainty in 
disk fractions around very low-mass objects. Given the lower luminosities of M-stars 
and BD, the SEDs can look flattened or transitional after a little evolution in 
the innermost 0.1-1 AU region, even if the rest of the disk is flared 
and optically thick. 

\begin{figure}
\plottwo{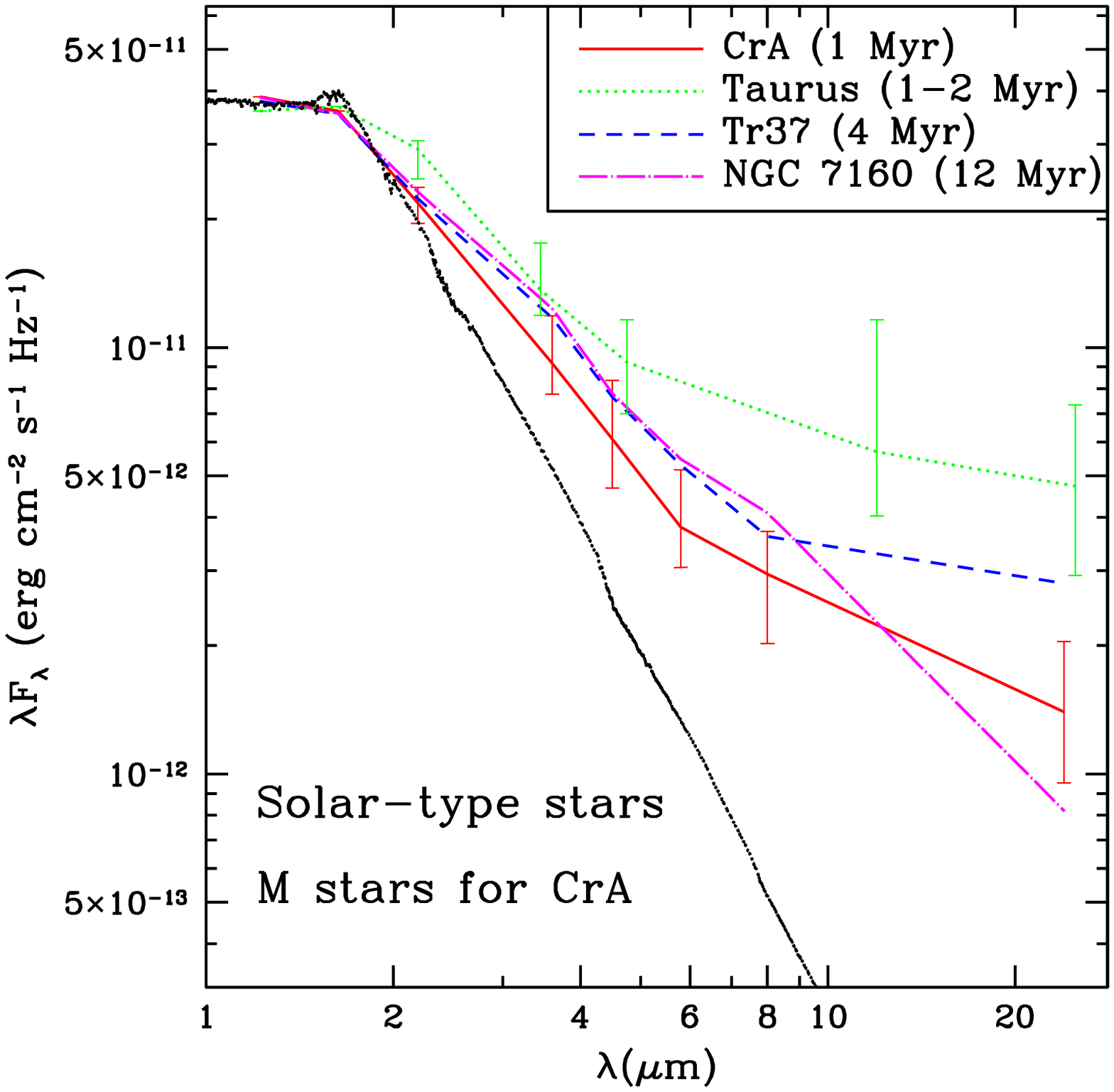}{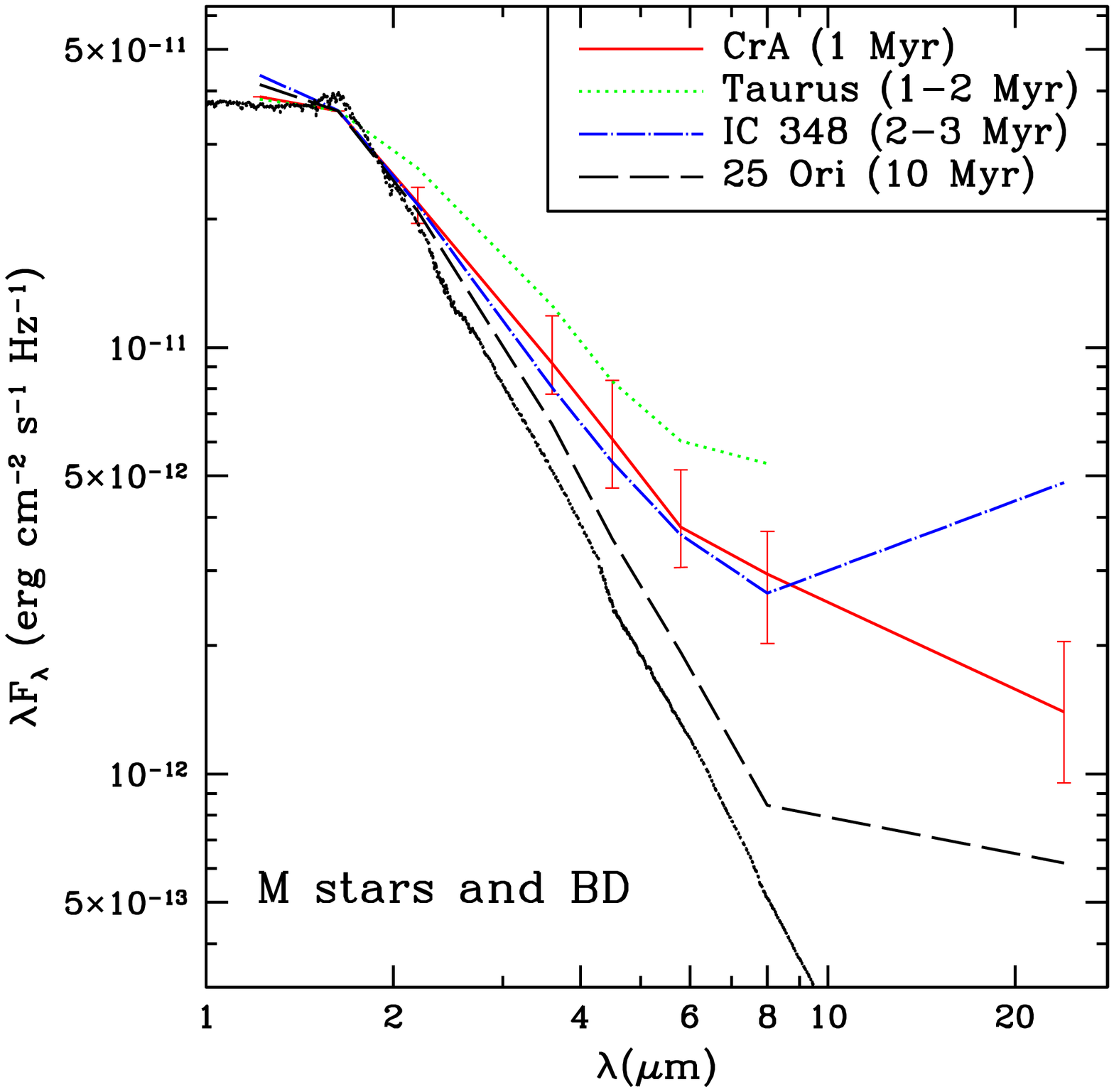}
\caption{Left: Median SED of the disks in the Coronet cluster, compared to 
the median SED of solar-type stars in regions at different stages of evolution 
(Taurus, Tr 37 , NGC 7160) and the photospheric emission of a K7 star. All the 
$\lambda$F$_\lambda$ fluxes have been scaled to the distance of the Coronet cluster 
(d$\sim$ 150 pc). The disks
around very low-mass objects in the CrA cloud have significantly less
near-IR emission than young disks in Taurus, and even less than the disks
in the very evolved NGC 7160. This suggest differences in disk structure for
M-type objects compared to solar-type stars. The quartiles are
shown for CrA and Taurus. For clarity, the quartiles of Tr 37 and NGC 7160
(which are similar to those of CrA) are not displayed.
Right: Median SED of the disks in the Coronet cluster, compared to the
median SED of M0-M7 objects in regions with different ages (Taurus, IC 348, 25 Ori).
All the fluxes have been scaled to the distance of the Coronet cluster(d$\sim$ 150 pc).
The typical disk SED of M-type objects at short wavelengths is close to a flat
disk at nearly all stages (25 Ori has only 4 disks, 3 of them TO). For clarity, the quartiles
are displayed for CrA only, being similar in the other regions.
\label{mediansed-fig}}
\end{figure}

A comparison of the median disk SED of different well-known regions and the Coronet
Cluster (Figure \ref{mediansed-fig} left) is a good indicator of the differences 
in disk morphology between M-type and solar-type stars. The median SED shows that 
the typical disks in the Coronet cluster are very different from the disks around 
solar-type stars in regions with similar ages (Taurus; Kenyon \& Hartmann 1995; 
Hartmann et al. 2005) and more similar to older regions (Tr 37 and NGC 7160; 
Sicilia-Aguilar et al. 2006a). Most of the Class II SEDs in the Coronet cluster have 
slopes close to flat disks ($\lambda$F$_\lambda$ $\sim$ $\lambda^{-4/3}$, optically
thick but geometrically thin disk; Hartmann 1998), which is very different from 
the SEDs of embedded clusters with similar age (e.g., the Tr 37 globule, 
Sicilia-Aguilar et al. 2006a). Nevertheless, the median SEDs for the older clusters 
in Figure \ref{mediansed-fig} (left) were calculated for solar-type stars, with spectral
types G to early M, very different from the M0-M8 spectral types in the
Coronet cluster. If the radial distance traced by the IR fluxes, and/or the
disk morphology, and/or their evolution vary with the
mass or spectral type of the star, comparing objects with different masses may
mask the real time evolution with age, or other effects such as influence of the
environment or initial conditions.

To explore the SED morphology of other very low-mass stars and BD, we trace the 
median disk SED for objects with M0-M8 spectral types in several regions for which accurate 
spectral types, extinctions, and Spitzer data were available: Taurus (23 objects, 
median spectral type M3, 1-2 Myr; Kenyon \& Hartmann 1995; Brice\~{n}o et al. 1998; 
Hartmann et al. 2005), IC 348 (128 objects, median spectral type M4,  2-3 Myr; 
Luhman et al. 2003; Lada et al. 2006);  25 Ori (4 objects, median spectral type M3, 
10 Myr; Brice\~{n}o et al. 2007; Hern\'{a}ndez et al. 2007b). The median SED was 
computed after correcting for extinction for each object. For Taurus and 25 Ori, we 
used the average A$_V$=1 mag and A$_V$=0.29 mag in the cases where the individual
extinction was not known. Since the typical extinctions are low in both regions, 
individual variations are not significant at IR wavelengths. As in the case of the 
Coronet cluster, we did not exclude TO disks, if their excesses at mid-IR were comparable 
to normal, optically thick disks (flared or flat). The median SEDs are displayed in 
Figure \ref{mediansed-fig} (right). All the regions, except Taurus, show a median SED 
at $\lambda <$6\,$\mu$m consistent with an optically thick, geometrically thin disk 
($\lambda$ F$_\lambda \sim \lambda^{-4/3}$; Hartmann 1998). For 25 Ori, 3 of the disks 
are TO with no excess at  $\lambda <$8\,$\mu$m, which explains the photospheric fluxes
down to this wavelength. The comparison with the median SED for solar-type stars in
these (Lada et al. 2006) and the regions in Figure \ref{mediansed-fig} (left) confirms
the differences in SED morphology depending on the spectral type of the central 
object.

The main conclusion from the median SED analysis is that flattened and transitional 
disks are far more common among the low-mass objects than among solar-type stars at
any given age and in most regions, extending the results of Megeath et al. (2005) 
and McCabe et al. (2006). One reason for this behavior could be that
a given IR wavelength traces a zone much closer to the star in a mid- or late-M
star than in a solar-type star, so any signs of disk evolution (grain growth,
disk flattening, inner holes) affecting the innermost disk ($\sim$0.1-1 AU) will 
produce larger changes in the near-IR flux for the low-mass objects than for
solar-type stars. On the other hand, 
the fact that the median SED of the young Coronet cluster is very close to a flat disk 
suggests that differences in the inner disk structure (maybe caused by lower disk masses 
and the lack of dead zones; Hartmann et al. 2006) and/or differences in the initial
conditions at the time of formation (for instance, in the angular momentum of the 
cores that form very low-mass stars; Dullemond et al. 2006) may shape the disks 
already at a very early stage. In addition, 
the environment (high-mass versus low-mass star forming regions; Luhman et al. 2008) 
could affect both the formation and the evolution of the objects and their disks, so 
future observations of low-mass objects in high-mass star forming regions 
should be performed to compare their properties with those in relatively isolated,
low-mass star-forming regions.

The large number of TO needs also to be considered. Typically, TO represent 
only 5-10\% of the total number of disks around solar-type stars (Sicilia-Aguilar 
et al. 2006a; Hartmann et al. 2005), which suggests that their lifetimes are less 
than a tenth of the lifetime of the optically thick disks, of the order of 
10$^5$-10$^6$ years. The situation is similar in young (e.g., Taurus, 1-2 Myr;
Hartmann et al. 2005) and intermediate-age (e.g. Tr 37, 4 Myr; Sicilia-Aguilar et al. 2006a) 
clusters, so disks with inner holes are thought to be on a rapid, transitional phase 
between Class II and Class III objects, lasting less than 1 Myr. The rapid 
transitional phase is usually interpreted as planetesimal or planet formation 
(Quillen et al. 2004), and/or strong dust coagulation to pebble-sized grains (starting
in the inner disk and progressing rapidly to the rest of the disk), and/or 
the effect of photoevaporation by the ultraviolet radiation of the central star 
(Alexander et al. 2006a,b). Timescales in the inner disk are shorter, so the 
evolution would occur faster in the innermost part (Hayashi et al. 1985), 
and extend rapidly to the rest. In certain cases, the inner holes may be also produced
by close-in companions (e.g. CoKu Tau/4; Ireland \& Kraus 2008). 

The flux excesses over the photosphere of most TO (Table \ref{excess-table}) are 
consistent with optically thick disks with inner holes rather than with debris disks 
(Kenyon \& Bromley 2005), except for the small excesses around some of the 
Class III sources. Therefore, they are most likely protoplanetary disks with inner 
holes (``transition objects'') than evolved, reprocessed, debris disks, although most 
debris disks studies have been done for intermediate-mass and solar-type stars.
The large fraction of TO (50\%) in the Coronet cluster points to a much 
longer ``transitional'' phase, with lifetimes of the same order of magnitude 
as the typical disk lifetimes. This is in contradiction with clearing up of inner 
holes via photoevaporation, which would remove the remaining disk in 10$^5$-10$^6$ 
years after the innermost disk has been photoevaporated (Alexander et al. 2006a,b). 
Observations of M-type stars and BD with disks have suggested that TO and 
disks with inner holes are more frequent around them (Mohanty et al. 2004; 
Allers et al. 2006; McCabe et al. 2006; Bouy et al. 2007), but so far the number of 
objects considered was significantly smaller than in the Coronet cluster. As with
the median SED analysis, the youth of the Coronet cluster suggests that some of the 
disks could be already formed as ``flattened'' disks or ``transition-like'' structures. 
More observations in different regions are needed to explore whether the differences 
in disk structure are related to evolution around a very low-mass object, and/or to 
the formation of very low-mass objects with disks, and/or the environment
in a low-mass star-forming cluster.

Finally, 5 objects are consistent with very embedded Class II or Class I sources 
according to their IRAC and MIPS colors (G-36, G-43, G-45, G-112, and G-122; see 
Figure \ref{sed-fig} and notes to Table \ref{obs-table}), and 20 more may be very 
embedded cluster members (mostly Class I, maybe Class 0 objects), with incomplete 
IRAC/MIPS data (due to low S/N or saturation by nearby extended sources). None of 
the very embedded sources nor potential embedded sources is associated with the 
millimeter clumps detected in the region (Henning et al. 1994; Chini et al. 2003), 
probably because of the low mass of the objects, as the surveys were mostly sensitive 
to masses typical of HAeBe and intermediate-mass TTS. In addition, the SEDs 
of several sources (see Table \ref{obs-table}) and their positions within
the cloud suggest that they are probably contaminating extragalactic objects.
For one of them, G-115, the optical spectrum confirms that it is a quasar
at redshift 0.62. The rest of them were not detected at optical wavelengths.

\subsection{Accretion in the Coronet cluster\label{accretion}}

\begin{figure}
\plotone{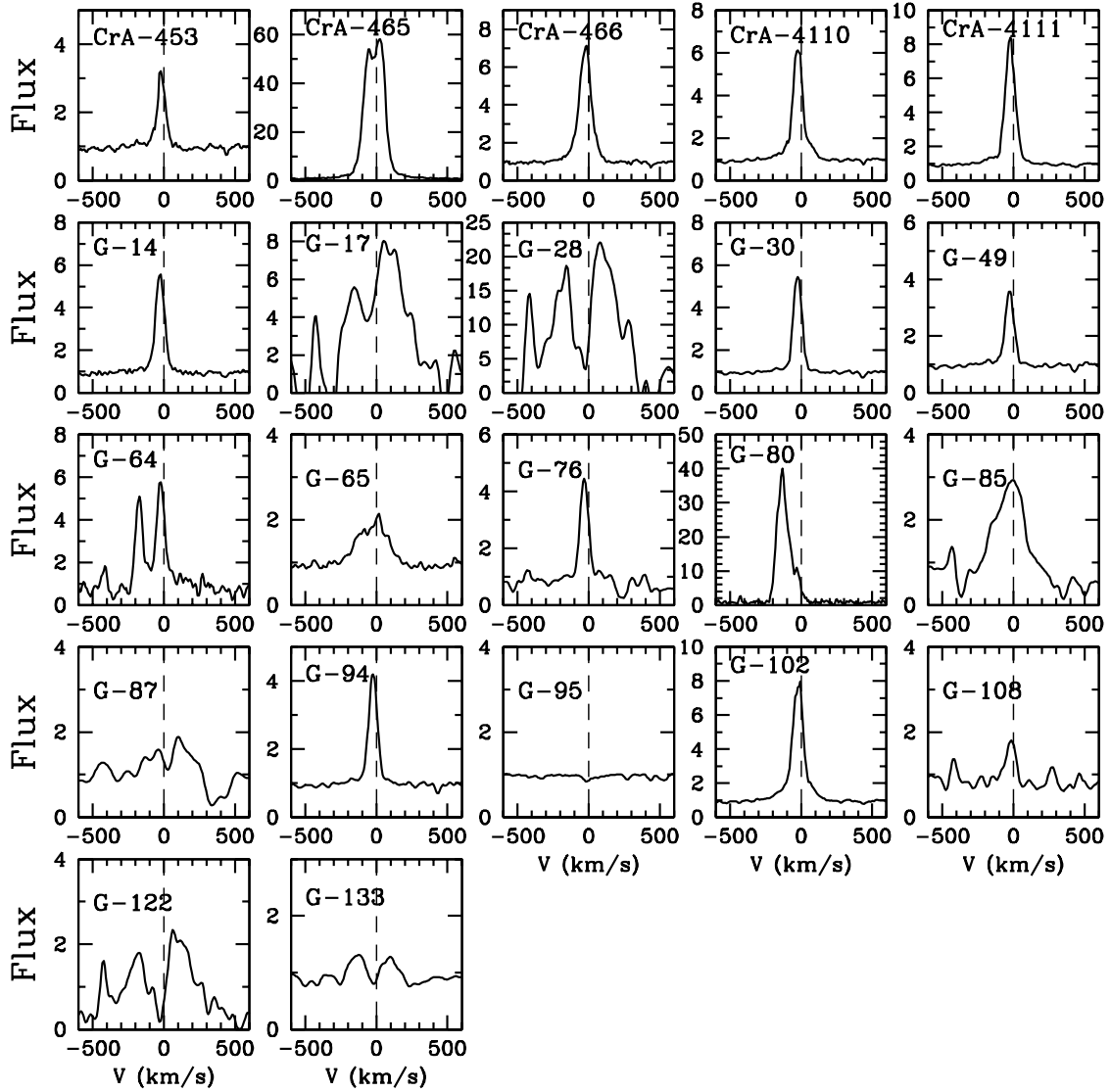}
\caption{H$\alpha$ emission of the very low-mass objects in the Coronet cluster.
The flux is normalized to the continuum level.
Most of the objects display the narrow lines typical of non-accreting systems.
The Class III object G-95 does not show any H$\alpha$ emission, but its youth
and membership are confirmed by the Li I absorption.
The detections of G-87 and G-133 are marginal. \label{halpha-fig}}
\end{figure}

The optical spectroscopy can be used to estimate the fraction of accreting disks 
by measuring the H$\alpha$ equivalent width (EW, see Table \ref{lines-table}, 
Figure \ref{halpha-fig}). EW larger than 10\,\AA\ or 20\,\AA\ in emission indicates
active accretion for objects with spectral types K3-M2 and later than M2,
respectively (White \& Basri 2003; a more precise classification is given by
Barrado y Navascu\'{e}s \& Mart\'{\i}n 2003). The H$\alpha$ emission lines of
accreting stars are usually broadened by the accretion and winds, so the
line profiles and the velocity wings are an
additional criterion to detect accretion (Muzerolle et al. 1998a, 2005; 
Natta et al. 2005), although the very low-mass objects tend to have narrower
H$\alpha$ profiles because of their smaller gravitational potential (Muzerolle
et al. 2005). The H$\alpha$ EW criterion is sometimes unable to detect 
accretion in objects with very low accretion rates ($<$10$^{-9}$\,M$_\odot$/yr; 
Sicilia-Aguilar et al. 2006b), but the intermediate resolution of FLAMES reveals 
H$\alpha$ line broadening larger than $\sim$120 km/s. The Ca II IR triplet 
($\lambda \lambda$ 8498, 8542, 8662\,\AA) is a marker of strongly accreting objects, 
and is included in the spectral range covered with FLAMES, but it is not detected 
in any of the sources except for a marginal detection in G-85 ($\sim$3-4$\sigma$,
with EW -0.6, -1.3, and -0.8 for the lines $\lambda\lambda$ 8498, 8542, and 8662\,\AA,
respectively). This suggests that the accretion rates in the Coronet cluster
are low to moderate ($<$10$^{-8}$-10$^{-9}$\,M$_\odot$/yr), as is typically observed 
for very low-mass stars and BD (Natta et al. 2004).

H$\alpha$ emission is detected in 21 objects (Figure \ref{halpha-fig}), although
the detection is marginal in G-87 and G-133. The star G-95 is a rare case, as it 
does not show any H$\alpha$ emission but a small absorption, which is very unusual 
even for a M1-type Class III object, since young M stars have very active chromospheres. 
Inclination and, to a lesser extent, rotation, may affect the line profiles and width 
(Muzerolle et al. 1998b, 2001), but the presence of absorption instead of emission is not
easy to reproduce by just adjusting these parameters. Nevertheless, the status of G-95 
as young star (and thus, cluster membership) can be inferred from the presence of Li I 
absorption. The results are that 10 out of 14 stars are weak-lined T Tauri stars (WTTS) 
class, with small H$\alpha$ EW and no signs of line broadening  nor other accretion 
indicators. This gives an accretion fraction of $\sim$30\%, which is lower than observed 
even in more evolved regions (Sicilia-Aguilar et al. 2006b). Nevertheless, some of these 
sources could be accreting at very low rates ($\sim$10$^{-10}$-10$^{-12}$\,M$_\odot$/yr), 
which produce small EW and no significant line broadening for the resolution of FLAMES 
(Muzerolle et al. 2005). Such low accretion rates are relatively normal for low-mass objects 
(Natta et al. 2004; Luhman et al. 2007). Among the TO objects, 
CrA-466 has strong EW indicative of accretion, and G-65 shows broad H$\alpha$, despite the 
small EW, so it is likely to be accreting at a low rate. All the Class II sources with near-IR 
excess are accreting, except for CrA-4110, which shows narrow H$\alpha$ with small EW (it 
could be accreting at a low rate, $\sim$10$^{-10}$-10$^{-12}$\,M$_\odot$/yr). None of the 
Class III objects shows any indications of accretion. If we extrapolate those results to
the total sample of Class III, Class II and TO, we obtain an accretion fraction $\sim$40\%, 
which is lower than observed for solar-type stars in similarly young regions. This is 
consistent with the observations in other regions (Muzerolle et al. 2003; 
Mohanty et al. 2005; Bouy et al. 2007), which also revealed a low fraction of accreting 
disks for the very low-mass members. 

We find the same relation between accretion rates and SED morphology than in 
the older cluster Tr 37 (Sicilia-Aguilar et al. 2006b): Some of the TO have ongoing 
accretion despite their lack of near-IR excess, while some normal-looking Class II 
sources have very low accretion rates despite not having inner holes. Therefore, we once
more observe that the ``classical'' distinction between CTTS and WTTS based on the
H$\alpha$ EW does not always correspond to Class II and Class III objects according to
their IR excesses (Hillenbrand 2008). Not only some objects with low accretion rates
show ``weak" (but broad) H$\alpha$ emission (Sicilia-Aguilar et al. 2005, 2006b), but
we find accreting and non-accreting objects among the disks with inner holes, and disks 
without holes but with very low accretion rates.

\subsection{X-ray properties of the very low-mass objects \label{xray}}

The temporal X-ray  properties of  each source were investigated by examining the arrival 
times of the photons from a given source using the software tool Event 
Browser\footnote{http://www.astro.psu.edu/xray/docs}.  The mean arrival rate of photons 
for each source was computed. If the arrival rate of events exceeded this expected rate 
by more than 1.3$\sigma$ as determined by Poisson probability of such an excess (Gehrels 1986), 
then the source was flagged as a flaring source and the length of the excess flux was 
determined as well as the Poisson probability of the excess.  Sources G-28, G-45, G-78 
and G-108 are only visible as flare sources, there was no steady flux. These results are 
shown in Table \ref{xray-table}. The complete census of X-ray sources and their properties
will be published in a separate work.

The hardness ratio (HR) and its error were computed for each source. The HR is defined as 
HR=(C(2-8)-C(0.5-2))/C(0.5-8), where C(x-y) is the background subtracted flux in the x to y 
keV energy band and the error for low count sources was computed using the method of Lyons (1991). 
The absorption column to the source, N$_H$, was either determined by a fit to the spectrum using the 
XSPEC software (Arnaud 1996) for sources with more than 100 photons or by computing the average 
energy of the lowest two photon energies (E$_{2ave}$) from the source for source with 3 - 100 photons.  
The computation of N$_H$ was made using a power-law spectrum with a photon index of -2 and a series of 
trial values of N$_H$. Other spectral forms such as bremsstrahlung and plasma models provide almost 
identical curves, implying that this method is fairly insensitive to spectral shape. The two lowest energy 
photons formed a fraction of the total flux which could be read off of the model curves. The E$_{2ave}$ 
in the source spectrum was then used to determined which N$_H$ best matched this energy.  The 
determination of N$_H$ upper and lower bounds was calibrated against photons selected at random 
from bright sources where the N$_H$ was determined by spectral fitting in XSPEC.  Distributions of 
the lowest two energy counts were created for sources with 3, 6, 10, 20, and 40 photons to estimate 
the errors using this method.  The results were tabulated and a linear regression fit was made to 
E$_{2ave}$ , and to the upper N$_H$ upper and lower bounds to the E$_{2ave}$  from the distributions as a 
function of source intensity and E$_{2ave}$. A more complete description of this method is being 
prepared for publication. The errors in Table \ref{xray-table} are the 95\% errors.  The luminosities were based 
on the unabsorbed flux either computed using the observed flux, absorption and an assumed spectrum 
of a 2 keV plasma and the PIMM's Tool\footnote{http://heasarc.gsfc.nasa.gov/Tools/w3pimms.html}, 
or from XSPEC by freezing all the parameters and setting 
the absorption to zero.  The distance was taken as 130 pc for the estimates.

The column density N$_H$ can be used to estimate the extinction independently from
the results derived from the colors and spectral types. Using the results of
Vuong et al. (2003), that N$_{H}$/A$_J$=5.6$\pm$0.4 10$^{21}$ cm$^{_2}$ mag$^{-1}$,
and the conversion A$_J$/A$_V$=0.282 in Cardelli et al. (1989), we obtain A$_V$
(see Table \ref{xray-table}). These values roughly agree with the values derived
previously, although the predicted extinctions are lower for some of the most extincted
objects. Since the standard relations between N$_H$ and A$_V$ are derived for the
interstellar medium, differences in dust and gas in the circumstellar environment
may be responsible for this effect, as well as the larger uncertainties in the derivation
of N$_H$ from X-ray data. In any case, we prefer the results derived from
spectral types and JHK photometry, which are in very good agreement with the fitted
SEDs.

\begin{figure}
\plotone{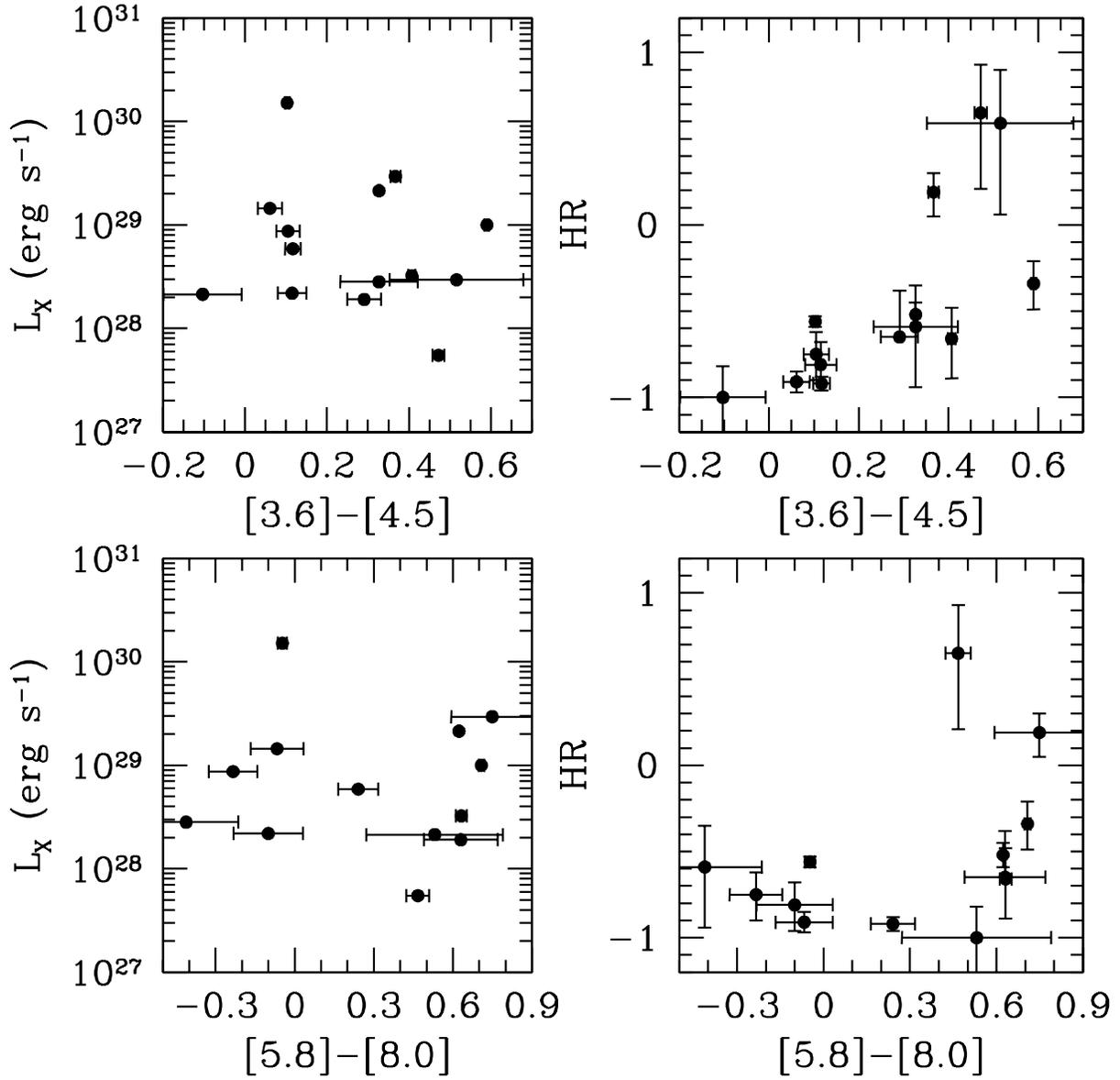}
\caption{X-ray luminosity and hardness ratio (HR) versus IRAC colors [3.6]-[4.5] 
and [5.8]-[8.0] (not corrected for reddening). There is a trend of objects with redder disks 
to have lower X-ray luminosities, and more positive HR, which is directly related 
to the extinction (the two objects in the upper right corner are the Class I source
G-122 and the very extincted TO G-87). \label{xray-fig}}
\end{figure}

Most of the Class II and Class III sources show negative HRs corresponding to 
moderate absorption and temperature around 2 - 10 keV.  The TO and Class I sources on the other
hand show a mixture of both negative and positive HRs. There is a correlation between the X-ray
luminosity and HR and the IR colors (Figure \ref{xray-fig}). Sources with smaller IR excess
(Class III and TO) tend to show higher X-ray luminosity and specially, more negative HR.
This correlation is likely caused by extinction effects, and could also suggest
differences in the orientation of the disks (closer to edge-on if the X-ray absorption is
stronger). Part of the difference in X-ray luminosity may be caused by the presence of close-in
binaries, which have also a noticeable effect on disks (Bouwman et al. 2006). So far
there is no information about binarity among the Coronet cluster members, but the holes in
some of the TO could be produced by close-in binaries (Ireland \& Kraus 2008). The Class I sources 
also display a large range of N$_H$, possibly because these sources are deeper in the cloud or 
their disks/envelopes are adding some absorption.  Many of the X-ray sources show 
considerable variability over longer time intervals. Seven additional observations were made 
of this region with some sources fading below detectability and others brightening.

\subsection{HH objects with X-ray emission in the Coronet cluster \label{environment}}

The whole CrA region is characterized by an extended, cometary-like nebulosity, which 
becomes very optically thick near the star R CrA (Graham 1992; Henning et al. 1994; 
Cambr\'{e}sy 1999). There are numerous Class I and Class 0 objects within the cloud 
(Henning et al. 1994; Chen et al. 1997), and a large number of shocks and HH objects, 
mostly detected via their optical emission (Wang et al. 2004). Some HH objects are known 
to produce X-ray emission (Pravdo et al. 2001; Favata et al. 2002; Raga et al. 2002), although
previous studies of X-ray emission in the Coronet cluster (Forbrich \& Preibisch 2007)
had not found any relation between the known HH objects and the X-ray sources.

\begin{figure}
\plotone{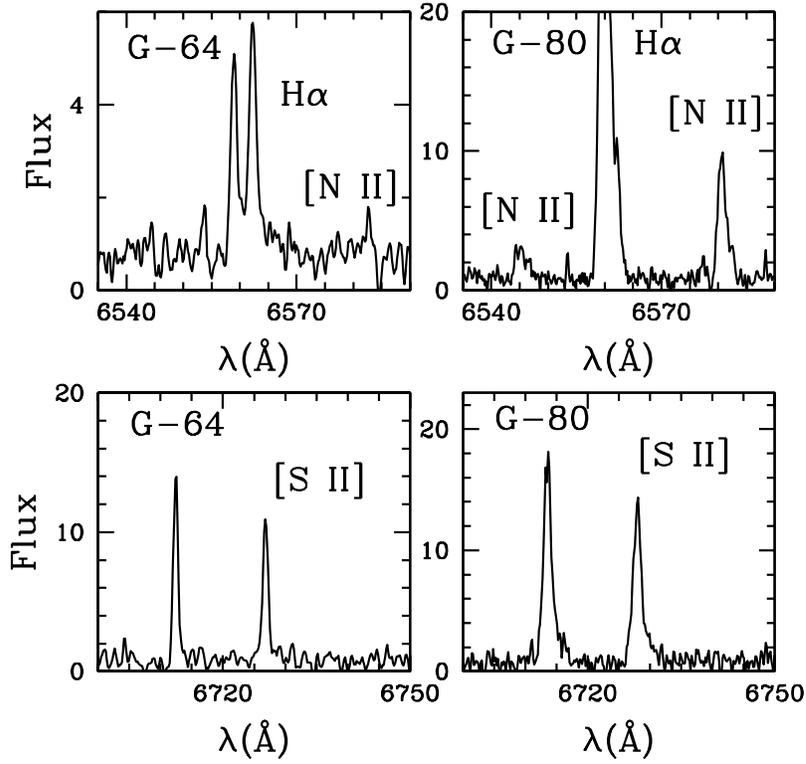}
\caption{Shock [N II] and [S II] lines observed in the X-ray emitting HH objects G-64 and G-80.
The H$\alpha$ line of G-80 shows the typical profile of shocks as well. \label{lines-fig}}
\end{figure}

Using the optical spectra, we found that two of the X-ray detected 
objects (G-64 and G-80) show the forbidden [S II] emission lines 
at 6716 and 6731\,\AA, characteristic of shocked material, as well as [N II] 
emission at 6548 and 6583\,\AA\  (Figure \ref{lines-fig}; Hamann 1994). 
In the case of G-64, Spitzer data reveals a nearby faint source, 
probably an embedded Class I source. G-64 is also located near a known
HH object, HH 96 (19:01:41.87 -37:00:57.3; Wang et al. 2004). 
On the other hand, G-80 is not detected at any of the Spitzer wavelengths,
being probably a HH knot in the cloud. Its H$\alpha$ profile, as well
as the [N II] and [S II] profiles, show the red tails characteristics of 
HH objects. G-80 is relatively far from the other known HH objects, and given 
the large number of sources in its surroundings, it is not easy to identify 
the object that produces it.

The EW of the lines are listed in Table \ref{shock-table}.
The line ratios are thus 0.84 and 1.22 for G-64 and G-80, respectively.
The line ratios depend on the electron density and (weakly) on the temperature
(Osterbrock 1989). Assuming a shock temperature of 10,000 K, the ratios
observed in the G-64 and G-80 suggest electron densities of the order of
$\sim$4$\times$10$^2$ cm$^{-3}$ and $\sim$10$^{3}$ cm$^{-3}$, respectively.
These values are slightly lower than the typical electron densities observed
in shocks near TTS (Hamann 1994), which tend to be close to saturation
($\sim$5$\times$10$^3$ cm$^{-3}$). Given the weak dependence of the 
electron density on the temperature as T$^{-1/2}$, temperature variations cannot
account for this difference, so we may conclude that electron densities in 
the CrA cloud are lower than the typical values.

\section{Conclusions \label{conclusions}}

The study of the disks around very low-mass objects in the Coronet cluster 
reveals significant differences with other regions. The disk fraction is 
smaller ($\sim$70\%) than in clusters with a similar age, and more striking, 
about $\sim$50\% of the disks show ``transitional'' SEDs with inner holes (TO). The SEDs 
of the disks without inner holes are significantly flatter than observed in 
similarly young clusters and even in regions a few Myr older 
(Sicilia-Aguilar et al. 2006a), and the median SED is close to that of a flat disk 
($\lambda$F$_\lambda$ $\sim$ $\lambda^{-4/3}$). The disks with no inner gaps
show typically evidence of accretion, while only some of the TO are actively
accreting. In any case, the H$\alpha$ EW and profiles suggest that accretion rates
are probably low to moderate ($<$10$^{-8}$-10$^{-9}$\,M$_\odot$/yr), as it is
common for very low-mass objects.

Since the members of the Coronet cluster have later spectral types than 
typical disks studies (M0-M7.5 instead of G-early M), our work suggests
that M-type stars and BD tend to have flatter disks and inner holes with a 
significantly higher frequency than solar-type stars. Partial evidence
of this had been suggested in other studies, based on smaller samples
of stars (Megeath et al. 2005; McCabe et al. 2006; Lada et al. 2006; 
Allers et al. 2006; Morrow et al. 2007). 

The fact that most disks around the very low-mass members of the Coronet 
cluster have inner holes or are strongly flattened may be responsible 
for the controversy about disk fractions for very low-mass stars and BD.
It also reveals that observations at mid-IR wavelengths are required in order
to determine accurately the disk fractions for M-type stars and BD.

The large fraction of ``transitional'' disks suggest a lifetime comparable 
to the typical disk lifetimes, which is inconsistent with the rapid timescales 
due to disk photoevaporation (Alexander et al. 2006a,b), and with the relatively 
short lifetimes inferred for solar-type stars in other regions (Hartmann et al. 2005; 
Sicilia-Aguilar et al. 2006a). Therefore, at least some of the ``transitional'' 
disks may not be in a short-lived, intermediate evolutive stage.

The youth of the Coronet cluster members suggests that part of these disks 
could have been formed already as flattened or ``transition-like'' structures. 
Whether this effect is due to the formation conditions of these low-mass objects 
in a low-mass star-forming region like the Coronet cluster (angular momentum of 
the collapsing cores, presence of binary companions), or is a consequence of a 
different evolutive path of disks around the less massive sources, needs 
to be explored by targeting large samples of very low-mass objects in 
older regions and in massive star-formation environments.

Finally, the optical spectroscopy confirms that one of the X-ray sources 
in the Coronet cluster corresponds to a newly detected HH object.

\acknowledgements
We want to thank A. Mart\'{\i}nez-Sansigre for his help identifying extragalactic 
sources, and the anonymous referee for his/her comments and the rapid response. 
A. \& G. Garmire acknowledge financial support from NASA grants SV4-74018, NAS8-01128, and
NAS8-38252. This work is based on observations made with the Spitzer Space Telescope, 
which is operated
by the Jet Propulsion Laboratory, California Institute of Technology under a contract with 
NASA. It also makes use of data products from the Two Micron All Sky Survey, which is 
a joint project of the University of Massachusetts and the Infrared Processing and Analysis 
Center/California Institute of Technology, funded by the National Aeronautics and Space 
Administration and the National Science Foundation, and the NIST Atomic Spectra Database
(http://physics.nist.gov/asd3).

\clearpage

\begin{deluxetable}{llllllll}
\tabletypesize{\scriptsize}
\tablenum{1}
\tablecolumns{8} 
\tablewidth{0pc} 
\tablecaption{Observations summary\label{obs-table}} 
\tablehead{
 \colhead{Name} & \colhead{Optical} & \colhead{IRS} & \colhead{2MASS} & \colhead{IRAC} & \colhead{MIPS} & \colhead{Chandra}  & \colhead{Class}} 
\startdata
CrA-205  & ---        & 2400-1680-1200-1200 & Det. & Det. & Det.  & --- & TO,4  \\
CrA-432  & ---        & 960-420-600-720 &  Det. & Det. & Det.  & --- & II,4 \\
CrA-453	 & Det.   & --- &  Det. & Det. & Det.  & --- & III, 2,4  \\
CrA-465	 & Det.   & 960-960-480-480 &  Det. & Det. & Nebula  & --- & II,4  \\
CrA-466	 & Det.  &  180-180-90-90  &  Det. & Det. & Det.  & Det. (G-113)  & TO, 1, 2,4 \\
CrA-4107  & ---        & 180-180-600-600 &  Det. & Det. & Det.  & --- & II,4  \\
CrA-4109  & ---        & 180-180-600-600 &  Det. & Det. & Det.  & ---  &  TO,4  \\
CrA-4110 & Det.   &  360-240-600-720 &  Det. & Det. & Det.  & ---  & II, flat, 2,4  \\
CrA-4111 & Det.   &  720-720-600-720 &  Det. & Det. & Det.  & Det. (G-110) &  TO, 1, 2,4 \\
G-1	 & ---   & 18-18-18-42 & Det. & Det. & Det.  & Det. & II, flat, 1, 2 \\
G-4	 & Non-det.     & --- & Non-det. & Faint & Non-det. & Det. & eg:, 1, 2 \\
G-6	 & Non-det.     & --- & Non-det. & Det. & Det.  & Det. & I:, emb, 1, 2 \\
G-10	 & Non-det.     & --- & Non-det. & Faint & Faint & Det. &  eg:, 1 \\
G-11	 & Very Faint   & --- &  Non-det. & Faint & Non-det. & Det. & emb \\
G-14	 & Det.   & 720-720-360-360 &  Det. & Det. & Det.  & Det. & TO, 2 \\
G-16	 & Very Faint   & --- &  Non-det. & Faint & Faint & Det. & eg:, 1, 2\\
G-17	 & Faint   & --- &   Non-det. & Faint & Non-det. & Det. & I, emb\\
G-20	 & Non-det.     & --- & Non-det. & Faint & Non-det. & Det. & emb \\   
G-22	 & Non-det.     & --- &  Non-det. & Faint & Non-det. & Det. & emb, 2 \\
G-28	 & Faint   & --- &   Non-det. & Faint & Near bright & Det. & I, emb, 1, 2 \\
G-29	 & Non-det.     & --- &  Non-det. & Faint & Faint & Det. & emb, 2\\
G-30	 & Det.   & --- & Det. & Det. & Det.  & Det. & III, 1, 2  \\
G-32	 & Very Faint   & --- & Faint & Det. & Det.  & Det.& II, emb,  2\\
G-36	 & Non-det.     & --- & Non-det. & Det. & Det.  & Det. & I, emb, ext:, 1, 2 \\
G-43	 & Non-det.     & --- & Non-det. & Faint & Det.  & Det. & I, emb, ext: \\
G-45	 & Very Faint   & --- & Non-det. & Faint & Faint  & Det. & I, emb, 1, 2 \\
G-48	 & Very Faint   & --- & Non-det. & Faint & Non-det.  & Det. & emb, 1 \\
G-49	 & Det.   & --- & Det. & Det. & Det.  & Det. & TO, 1, 2  \\
G-53	 & Faint   & --- & Non-det. & Faint & Non-det.  & Det. & emb, near RCrA, 1, 2\\
G-57	 & Non-det.     & --- & Non-det. & Faint & Near bright.  & Det. & emb, near RCrA, 1, 2 \\
G-61	 & Very Faint   & --- &Non-det. & Faint & Non-det.  & Det. & emb, near optical HH, 1, 2\\
G-64	 & Faint   & --- & Non-det. & Faint & Faint  & Det. & I:, HH, 1, 2 \\
G-65	 & Faint   & 56-42-90-90 & Det. & Det. & Det.  & Det. & TO, 1, 2 \\
G-71	 & Very Faint   & --- & Non-det. & Very Faint & Non-det.  & Det. & emb/eg:, 1 \\
G-74	 & Non-det.     & --- & Non-det. & Faint & Non-det.  & Det. &  emb, 1, 2 \\
G-76	 & Faint   & --- & Non-det. & Faint & Non-det.  & Det. & I/II:, emb, 1\\
G-80	 & Faint   & ---& Non-det. & Non-det. & Non-det. & Det. & HH \\
G-85	 & Faint   & 42-42-90-90 & Det. & Det. & Det.  & Det. & II, 1, 2 \\
G-87	 & Faint   & 180-180-90-90 & Det. & Det. & Det.  & Det. & TO, 2\\
G-88	 & Very Faint   & ---  & Non-det. & Faint & Non-det.  & Det. & emb, 1, 2 \\
G-90	 & Very Faint   & --- &Non-det. & Faint & Non-det.  & Det. & emb, 1 \\
G-94	 & Det.  & 240-240-360-360 & Det. & Det. & Det.  & Det. & TO/III, 1, 2 \\
G-95	 & Det.   & 42-42-90-90 & Det. & Det. & Det.  & Det. & TO , 1, 2\\
G-99	 & Non-det.     & --- & Non-det. & Faint & Extended  & Det. & emb, ext 24$\mu$m, 1, 2 \\
G-100	 & Non-det.   & --- & Non-det. & Faint & Non-det.  & Non-det. &  emb, ext, 3 \\
G-101	 & Very Faint:   & --- & Non-det. & Faint & Non-det.  & Det. & I: emb, 1 \\
G-102	 & Det.   & --- & Det. & Det. & Det.  & Det. & TO/III, 1, 2 \\
G-105	 & Very Faint   & --- & Non-det. & Faint & Faint  & Det. &  eg:\\
G-108	 & Very Faint   & --- & Non-det. & Faint & Faint  & Det.& I:/eg:, 1\\
G-109	 & Non-det.     & --- &Non-det. & Very Faint & Non-det.  & Det. & emb, 1 \\
G-112	 & Very Faint   & --- & Non-det. & Det. & Det.  & Det. & I, emb, ext, 1, 2 \\
G-114	 & Non-det.    & --- & Non-det. & Faint & Non-det.  & Det. & emb\\
G-115	 & Det.  & --- & Non-det. & Det. & Det.  & Det. & eg, quasar, Z=0.62, 1, 2 \\
G-116	 & Non-det.    & --- & Non-det. & Faint & Non-det.  & Det. & emb, 1\\
G-119	 & Non-det.     & --- & Non-det. & Very Faint & Non-det.  & Det. & eg:, 1\\
G-122	 & Faint   & --- &  Non-det. & Faint & Extended:  & Det. & I, ext\\
G-128	 & Very Faint:   & --- & Non-det. & Very Faint & Non-det.  & Det. & emb,1  \\
G-130	 & Non-det.     & --- & Non-det. & Faint & Non-det.  & Det. & near II, emb, 1 \\
G-132	 & Very Faint:   & --- & Non-det. & Faint & Near bright  & Det. & emb, near RCrA, 1 \\
G-133	 & Faint   & --- &  Non-det. & Very Faint & Near bright  & Non-det. & emb, near RCrA,3  \\
G-135	 & Very Faint:   & --- & Non-det. & Faint & Faint  & Det. & eg:, 1\\
G-136	 & Very Faint   & --- &  Non-det. & Very Faint & Non-det.  & Non-det. & eg:, 3 \\
\enddata
\tablecomments{Summary of observations. The `Optical' column indicates whether the object was 
observed with FLAMES. The `IRS' column indicates the exposure times (in seconds)
in each one of the four IRS modules for the objects observed with IRS.
The 2MASS, IRAC, MIPS, and Chandra columns indicate if the source was detected in
with these instruments. The following labels are used: ``Det.''= detected; ``Non-det.''=
not detected; ``Faint''= weak source; ``Very Faint'' = marginal detection; 
``Extended''= extended source; ``Nebula'' = source embedded in nebular emission;
``Near bright'' = source located near a very bright saturated source that does not allow
to extract the photometry.
The Class indicates the SED type (Lada et al. 1987): Class I (protostars or proto-BD), Class II 
(CTTS stars or BD analogs), TO (transition objects between Class II and Class III, with
optically thin inner disks), Class III (pre-main sequence stars or BD with no optically thick disks).
The class is determined via the IRAC/MIPS/IRS data and the H$\alpha$ line. Since all the
H$\alpha$ nebular emission in the sky positions within the cloud shows weak, narrow lines, 
objects with broad H$\alpha$ and uncertain IRAC/MIPS photometry are likely embedded Class I sources. 
``HH" denotes Herbig-Haro or shock objects. 
The note ``flat'' has been added when the disk (or transition disk) has a slope close to that of a
flat disk ($\lambda$F$_\lambda \propto \lambda^{-4/3}$. Embedded and/or heavily extincted
objects are labeled ``emb''. Some of them could be extragalactic sources located behind the
nebula. Extended objects or objects surrounded by nebulosity are labeled ``ext''. Extragalactic 
or possibly extragalactic sources are marked with ``eg''. Nearby source that may be affecting 
(contaminating or saturating) part or all of the
Spitzer data are listed. Colon ``:'' indicates uncertain values.
Extended sources, or sources surrounded by nebulosity, are labeled
``ext''. 1= X-ray variable source, according to Garmire \& Garmire (2003); 2= Listed
as X-ray source in Forbrich \& Preibisch (2007); 3= low-probability X-ray detection;
4= Source detected in optical (L\'{o}pez-Mart\'{\i} et al. 2005).
}
\end{deluxetable}

\begin{deluxetable}{lcccccccc}
\rotate
\tablenum{2}
\tablecolumns{9} 
\tablewidth{0pc} 
\tablecaption{X-ray properties\label{xray-table}} 
\tablehead{
 \colhead{Name} & \colhead{Class} & \colhead{Counts} & \colhead{Flare Time(s)} & \colhead{Significance} & \colhead{log(L$_X$ / erg s$^{-1}$)} & \colhead{HR} & \colhead{N$_H$ (10$^{22}$ cm$^{-2}$)} & \colhead{A$_V$ (mag)}} 
\startdata
CrA-466$^b$ & TO & 23 & 250 & 0.90 & 28.51 & -0.66$^{+0.18}_{-0.23}$ & 0.02$^{+0.38}_{-0.02}$  & 0.0$^{<0.2}_{>0.0}$ \\
CrA-4111$^a$ & TO & 17 & 833 & 0.90 & 28.33 & -1.00$^{+0.18}_{-0.20}$ & 0.03$^{+0.50}_{-0.03}$  & 0.0$^{<0.3}_{>0.0}$ \\
G-1 & II & 170 & 600 & 0.95 & 29.33 & -0.52$^{+0.07}_{-0.07}$ & 0.37$^{+0.71}_{-0.26}$  & 0.2$^{<0.5}_{>0.1}$ \\
G-6 & I & 11 & 487 & 0.90 & 28.03 & 0.60$^{+0.24}_{-0.35}$ & 0.02$^{+0.12}_{-0.02}$  & 0.0$^{<0.1}_{>0.0}$ \\
G-14 & TO & 15 & --- & --- & 28.28 & -0.65$^{+0.27}_{0.00}$ & 0.04$^{+0.54}_{-0.04}$  & 0.0$^{<0.3}_{>0.0}$ \\
G-17 & I & 6 & --- & --- & 30.14 & 1.00$^{+0.00}_{-0.23}$ & 61.33$^{+173.94}_{-48.72}$  & 31$^{<118}_{>6}$ \\
G-28* & I & 4 & 8,000 & 0.97 & 29.02 & 1.00$^{+0.00}_{-0.08}$ & 15.63$^{+38.38}_{-12.07}$  & 7.9$^{<27.1}_{>1.8}$ \\
G-30 & III & 84 & 240 & 0.90 & 29.16 & -0.91$^{+0.06}_{-0.06}$ & 0.18$^{+0.40}_{-0.18}$ & 0.1$^{<0.3}_{>0.0}$ \\
G-32 & II & 9 & 5,290 & 0.80 & 28.45 & -0.59$^{+0.24}_{-0.35}$ & 1.13$^{+3.52}_{-1.13}$  & 0.6$^{<2.3}_{>0.0}$ \\
G-36 & I & 47 & 2,600 & 0.90 & 29.50 & 0.69$^{+0.11}_{-0.13}$ & 1.56$^{+2.78}_{-1.25}$  & 0.8$^{<2.2}_{>0.2}$ \\
G-43 & I & 4 & --- & --- & 28.46 & 0.48$^{+0.35}_{-0.70}$ & 0.72$^{+2.80}_{-0.72}$  & 0.4$^{<1.8}_{>0.0}$ \\
G-45 & I & 25 & 1,100 & 0.90 & 29.18 & 0.64$^{+0.17}_{-0.21}$ & 3.58$^{+4.07}_{-2.98}$  & 1.8$^{<3.8}_{>0.3}$ \\
G-49 & III & 20 & 1,800 & 0.95 & 28.34 & -0.81$^{+0.13}_{-0.15}$ & 0.01$^{+0.26}_{-0.01}$  & 0.0$^{<0.1}_{>0.0}$ \\
G-64 & I & 11 & 6,000 & 0.84 & 28.63 & 0.27$^{+0.24}_{-0.37}$ & 1.34$^{+3.64}_{1.27}$  & 0.7$^{<2.5}_{>1.3}$ \\
G-65 & TO & 68 & 1,200 & 0.97 & 29.47 & 0.19$^{+0.11}_{-0.14}$ & 1.52$^{+2.55}_{-1.17}$  & 0.8$^{<2.0}_{>0.2}$ \\
G-76* & I & 4 & 8,400 & 0.97 & 28.94 & 0.49$^{+0.37}_{-0.79}$ & 2.17$^{+2.64}_{-2.17}$  & 1.1$^{<2.4}_{>0.0}$ \\
G-80 & HH & 5 & --- & 0.95 & 27.96 & -0.72$^{+0.36}_{-0.63}$ & 0.32$^{+1.59}_{-0.03}$  & 0.2$^{<1.1}_{>0.1}$\\
G-85 & II & 48 & 1,305 & 0.97 & 29.00 & -0.34$^{+0.13}_{-0.15}$ & 0.76$^{+1.18}_{-0.73}$  & 0.4$^{<1.0}_{>0.0}$ \\
G-87 & TO & 6 & --- & --- & 27.74 & 0.65$^{+0.28}_{-0.44}$ & 0.00$^{+0.64}_{0.00}$  & 0.0$^{<0.3}_{>0.0}$ \\
G-94 & III & 123 & 160 & 0.90 & 28.77 & -0.92$^{+0.04}_{-0.04}$ & 0.02$^{+0.37}_{-0.02}$  & 0.0$^{<0.2}_{>0.0}$ \\
G-95 & III & 789 & 30 & 0.97 & 30.18 & -0.56$^{+0.03}_{-0.03}$ & 0.65$^{+0.16}_{-0.17}$  & 0.3$^{<0.4}_{>0.2}$ \\
G-101 & I & 243 & 500 & 0.97 & 29.84 & -0.50$^{+0.06}_{-0.06}$ & 0.78$^{+0.73}_{-0.46}$  & 0.4$^{<0.8}_{>0.2}$ \\
G-102 & III & 59 & 2,000 & 0.97 & 28.94 & -0.75$^{+0.13}_{-0.15}$ & 0.03$^{+0.31}_{-0.03}$  & 0.0$^{<0.2}_{>0.0}$ \\
G-108* & I & 5 & 8,960 & 0.97 & 28.96 & 0.51$^{+0.38}_{-0.85}$ & 1.10$^{+3.00}_{-1.10}$  & 0.6$^{<2.1}_{>0.0}$ \\
G-112 & I & 16 & 1,235 & 0.99 & 28.52 & -0.23$^{+0.25}_{-0.37}$ & 0.41$^{+1.22}_{-0.41}$  & 0.2$^{<0.8}_{>0.0}$ \\
G-122 & I & 5 & --- & --- & 28.47 & 0.59$^{+0.31}_{-0.53}$ & 0.60$^{+3.13}_{-0.60}$  & 0.3$^{<1.9}_{>0.0}$ \\
G-130 & I & 8 & 205 & 0.90 & 28.00 & 0.39$^{+0.33}_{-0.64}$ & 0.00$^{+0.46}_{-0.00}$  & 0.0$^{<0.2}_{>0.0}$ \\
\enddata
\tablecomments{X-ray properties of the Class III, Class II, TO, and Class I objects and
the HH knot G-80. Sources marked with `*' were observed only as flare sources. Errors are 95\% errors.
$^a$ CrA-4111 corresponds to the X-ray source G-110. $^b$ CrA-466 corresponds to the X-ray source G-113.
The extinction A$_V$ is estimated from N$_H$ (Vuong et al. 2003; Cardelli et al. 1989)}
\end{deluxetable}

\clearpage
\pagestyle{empty}
\begin{deluxetable}{lcccccccccl}
\tabletypesize{\scriptsize}
\rotate
\tablenum{3}
\tablecolumns{11} 
\tablewidth{0pc} 
\tablecaption{2MASS and Spitzer (IRAC, MIPS) Data \label{archive-table}} 
\tablehead{
 \colhead{Name} & \colhead{ID} & \colhead{J}  & \colhead{H} & \colhead{K} & \colhead{3.6$\mu$m}& \colhead{4.5$\mu$m} & \colhead{5.8$\mu$m} & \colhead{8.0$\mu$m} & \colhead{24$\mu$m} & \colhead{70$\mu$m} } 
\startdata
CrA-205 & 19011169-3722213 & 13.315$\pm$0.023 & 12.741$\pm$0.023 & 12.405$\pm$0.023 & --- & 11.93$\pm$0.12 & --- & 12.42$\pm$0.38 & 9.10$\pm$0.14: &  --- \\
CrA-432 & 19005974-3647109 & 14.190$\pm$0.026 & 13.333$\pm$0.035 & 12.821$\pm$0.030 & 11.966$\pm$0.041 & 11.700$\pm$0.067 & 10.84$\pm$0.11 & 10.704$\pm$0.075 & 8.421$\pm$0.099 & --- \\
CrA-453 & 19010460-3701292 & 13.338$\pm$0.024 & 12.524$\pm$0.025 & 12.075$\pm$0.023 & 11.634$\pm$0.022 & 11.601$\pm$0.037 & 11.54$\pm$0.14 &  11.539$\pm$0.095 & --- &  --- \\
CrA-465 & 19015374-3700339 & 14.084$\pm$0.026 & 13.401$\pm$0.033 & 13.015$\pm$0.033 & 12.264$\pm$0.072 & 11.993$\pm$0.071 & 13.35$\pm$1.20: & 11.02$\pm$0.11 & 7.120$\pm$0.088 &  ---  \\
CrA-466 & 19011893-3658282 & 12.834$\pm$0.024 & 11.245$\pm$0.025 & 10.453$\pm$0.021 & 9.611$\pm$0.006 & 9.204$\pm$0.006 & 8.805$\pm$0.019 & 8.173$\pm$0.007 & 5.366$\pm$0.017 &  2.22$\pm$0.17 \\
CrA-4107 & 19025464-3646191 & 12.440$\pm$0.022 & 11.795$\pm$0.022 & 11.408$\pm$0.021 & 10.719$\pm$0.009 & --- & 10.043$\pm$0.035 & --- & 7.019$\pm$0.035 & 1.492$\pm$0.042 \\
CrA-4109 & 19021667-3645493 & 12.004$\pm$0.023 & 11.303$\pm$0.022 & 11.017$\pm$0.021 & 10.688$\pm$0.012 & --- & 10.601$\pm$0.082 & --- & 6.990$\pm$0.034 &  ---  \\
CrA-4110 & 19011629-3656282 & 12.954$\pm$0.021 & 12.314$\pm$0.022 & 11.897$\pm$0.021 & 11.208$\pm$0.015 & 10.846$\pm$0.023 & 10.777$\pm$0.067 &  10.191$\pm$0.031 & 7.410$\pm$0.085 &  --- \\
CrA-4111 & 19012083-3703027 & 13.233$\pm$0.024 & 12.687$\pm$0.023 & 12.402$\pm$0.024 & 11.950$\pm$0.031 & 12.059$\pm$0.063 & 11.71$\pm$0.19 & 11.193$\pm$0.079 & 6.63$\pm$0.16 &  --- \\
G-1 & 19022708-3658132 & 9.307$\pm$0.024 & 8.292$\pm$0.038 & 7.900$\pm$0.016 & 7.465$\pm$0.001 & 7.138$\pm$0.002 & 6.633$\pm$0.003 & 6.010$\pm$0.002 & 4.198$\pm$0.009 & 2.70$\pm$0.20  \\
G-6 & 19:02:21.9 -36:56:03 & --- & --- & --- & --- & 14.82$\pm$0.84: & 13.22$\pm$0.88: & 11.77$\pm$0.17 & 9.20$\pm$0.14 &  ---  \\
G-14 & 19021201-3703093 & 13.408$\pm$0.029 & 12.582$\pm$0.029 & 12.145$\pm$0.021 & 11.543$\pm$0.025 & 11.252$\pm$0.033 & 11.24$\pm$0.13 & 10.607$\pm$0.050 & 7.997$\pm$0.059 &  ---  \\
G-30 & 19020012-3702220 & 11.859$\pm$0.026 & 11.238$\pm$0.024 & 11.003$\pm$0.026 & 10.771$\pm$0.017 & 10.710$\pm$0.024 & 10.376$\pm$0.081 & 10.443$\pm$0.058 & 8.82$\pm$0.15 &   ---  \\
G-32 & 19015833-3700267 & 16.945$^a$ & 15.270$\pm$0.117 & 13.652$\pm$0.051 & 12.289$\pm$0.063 & 11.962$\pm$0.070 & 11.17$\pm$0.11 & 11.58$\pm$0.16: & 7.78$\pm$0.25 &  ---   \\
G-36 & 19:01:55.9 -36:52:05 & --- & --- & --- & 16.1$\pm$2.4 & 13.11$\pm$0.34 & 12.05$\pm$0.44 & 10.88$\pm$0.16 & 7.195$\pm$0.088 &   ---   \\
G-43 & 19:01:52.6 -37:02:44 & --- & --- & --- & 13.64$\pm$0.20 & --- & 11.22$\pm$0.17: & 9.323$\pm$0.025: & 10.0$\pm$0.5: &  ---  \\
G-45 & 19:01:52.2 -37:05:42 & --- & --- & --- & --- & --- & 14.3$\pm$2.2 & --- & --- &   1.89$\pm$0.15: \\
G-49 & 19014936-3700285 & 12.498$\pm$0.026 & 11.891$\pm$0.025 & 11.603$\pm$0.024 & 11.339$\pm$0.021 & 11.224$\pm$0.028 & 11.207$\pm$0.097 & 11.307$\pm$0.089 & --- &   ---  \\
G-65 & 19014041-3651422 & 13.898$\pm$0.032 & 11.629$\pm$0.023 & 10.481$\pm$0.019 & 9.403$\pm$0.009 & 9.036$\pm$0.008 & 8.395$\pm$0.067 & 7.65$\pm$0.14 & --- &  ---  \\
G-85 & 19013385-3657448 & 14.448$\pm$0.030 & 11.910$\pm$0.023 & 10.474$\pm$0.019 & 9.024$\pm$0.004 & 8.434$\pm$0.004 & 8.085$\pm$0.010 & 7.377$\pm$0.004 & 4.137$\pm$0.010 &   0.89$\pm$0.27 \\
G-87 & 19013232-3658030 & 14.853$\pm$0.037 & 12.622$\pm$0.029 & 11.432$\pm$0.023 & 10.352$\pm$0.009 & 9.880$\pm$0.011 & 9.629$\pm$0.040 & 9.162$\pm$0.016 & 6.291$\pm$0.056 &  ---  \\
G-94 & 19012901-3701484 & 11.637$\pm$0.024 & 10.956$\pm$0.023 & 10.663$\pm$0.021 & 10.386$\pm$0.010 & 10.269$\pm$0.016 & 10.278$\pm$0.067 & 10.037$\pm$0.037 & 8.76$\pm$0.21 &   ---  \\
G-95 & 19012872-3659317 & 10.828$\pm$0.023 & 9.591$\pm$0.025 & 9.002$\pm$0.019 & 8.709$\pm$0.003 & 8.606$\pm$0.005 & 8.429$\pm$0.013 & 8.476$\pm$0.010 & 7.70$\pm$0.12 &   ---  \\
G-100 & 19:01:25.8 -36:53:59 & --- & --- & --- & --- & --- & 11.21$\pm$0.20 & --- & --- &  ---  \\
G-102 & 19012562-3704535 & 12.363$\pm$0.023 & 11.654$\pm$0.021 & 11.299$\pm$0.023 & 10.861$\pm$0.010 & 10.800$\pm$0.018 & 10.534$\pm$0.055 & 10.596$\pm$0.043 & 9.37$\pm$0.49 &  ---   \\
G-112 & 19:01:19.4 -37:01:42 & --- & --- & --- & --- & 14.6$\pm$1.1 &  --- &  11.52$\pm$0.11 & 6.002$\pm$0.034 &  ---  \\
G-115 & 19:01:15.9 -37:03:44 & --- & --- & --- & 13.69$\pm$0.16 & 12.98$\pm$0.13 & 11.63$\pm$0.24 & 11.113$\pm$0.057 & 8.77$\pm$0.57  &  ---  \\
G-122 & 19:01:40.9 -36:57:15 & --- & --- & --- & 12.86$\pm$0.11: & 12.35$\pm$0.12: & 10.59$\pm$0.13: & 9.079$\pm$0.031: & 4.429$\pm$0.056:  & ---   \\
\enddata
\tablecomments{Archive data available through the 2MASS and Spitzer databases for the objects
observed with FLAMES and/or IRS. Only the objects detected clearly at 2 or more wavelengths are listed
here. ID indicates the 2MASS ID, or the coordinates, when 
the source was not detected by 2MASS. For the IRAC and MIPS photometry, the dominant source of error 
is the array location dependence, which results in typical 10\% errors. $^a$: Upper limit 
2MASS photometry. The colon (:) indicates uncertain values, mostly due to extended emission and/or
very low signal to noise ratio. }
\end{deluxetable}

\clearpage
\pagestyle{plaintop}

\clearpage
\pagestyle{empty}
\begin{deluxetable}{lcclll}
\tabletypesize{\scriptsize}
\rotate
\tablenum{4}
\tablecolumns{6} 
\tablewidth{0pc} 
\tablecaption{Spectral typing indices \label{index-table}} 
\tablehead{
 \colhead{Name} & \colhead{$\lambda$ Numerator} & \colhead{$\lambda$ Denominator} & \colhead{Range}   & \colhead{Calibration} & \colhead{Reference}} 
\startdata
PC1 & 7030-7050 & 7525-7550 &  M3-M9 & -0.06+2.95 X	 & 1,2 \\
PC2 & 7540-7580 & 7030-7050 &  M4-M8 & -0.63+3.89 X      & 1,2 \\
PC3 & 8235-8265 & 7540-7580 &  M3-M9 & -8.01+14.08 X-2.81 X$^2$ & 2 \\
PC4 & 9190-9225 & 7540-7580 &  M3-M9 & -0.94+4.66 X-0.52 X$^2$  & 2\\
R1  & 8025-8130 & 8015-8025 &  M2.5-M8 & 2.8078+21.085(X-1.044)-53.025(X-1.044)$^2$+60.755(X-1.044)$^3$ & 3 \\
R2  & 8415-8460 & 8460-8470 &  M3-M8 & 2.9091+10.503(X-1.035)-14.105(X-1.035)$^2$+8.5121(X-1.035)$^3$ & 3 \\
R3  & (8125-8130)+(8415-8460) & (8015-8025)+(8460-8470) & M2.5-M8 & 2.8379+19.708(X-1.035)-47.679(X-1.035)$^2$+52.531(X-1.035)$^3$ & 3 \\
TiO 8465 & 8405-8425 & 8455-8475 & M3-M8 & 3.2147+8.7311(X-1.085)-10.142(X-1.085)$^2$+5.6765(X-1.085)$^3$ & 3 \\
VO 2 & 7920-7960 & 8130-8150 & M3-M8 & 2.6102-7.9389(X-0.963)-8.3231(X-0.963)$^2$-14.660(X-0.963)$^3$ & 3 \\
VO 7445 & 0.5625(7350-7400)+0.4375(7510-7560) & 7420-7470 & M5-M8 & 5.0881+17.121(X-0.982)+13.078(X-0.982)$^2$ & 3 \\
\enddata
\tablecomments{Indices used for spectral typing. The wavelengths are given in \AA. 
In the calibration, X represents the index, and the resulting number indicates the
M subtype.
References: 1= Kirkpatrick et al. (1996); 2= Mart\'{\i}n et al. (1996); 3= Riddick et al. (2007).
The lineal relations for PC1 and PC2 are obtained by fitting together the objects in references
1 and 2.}
\end{deluxetable}

\clearpage
\pagestyle{plaintop}

\begin{deluxetable}{ll}
\tabletypesize{\scriptsize}
\rotate
\tablenum{5}
\tablecolumns{2}
\tablewidth{0pc}
\tablecaption{Parameters in the TLTD model\label{silparam-table}}
\tablehead{
 \colhead{Parameter} & \colhead{Meaning}}
\startdata
F$_\nu$	& Observed flux	\\
F$_{\nu,cont}$	& Total continuum flux	\\
d & Distance to the source, 150 pc\\
R$_*$ & Stellar radius, derived from spectral type and luminosity \\
T$_*$ & Stellar temperature, derived from spectral type \\
T$_{\rm a, min}$ & Lowest temperature in the disk atmosphere \\
T$_{\rm a, max}$ & Highest temperature in the disk atmosphere \\
T$_{\rm r, min}$ & Lowest temperature in the inner rim \\
T$_{\rm r, max}$ & Highest temperature in the inner rim \\
T$_{\rm m, min}$ & Lowest temperature in the disk midplane \\
T$_{\rm m, max}$ & Highest temperature in the disk midplane \\
D$_0$ & Scale of the contribution of the star (to allow 10\% error in luminosity) \\
D$_1$ & Contribution of the inner rim to the total flux \\
D$_2$ & Contribution of the midplane to the total flux \\
D$_{i,j}$	& Mass contribution of species i with size j \\
$\kappa _{i,j}$ & Mass absorption coefficient of species i with size j (Figure \ref{qval-fig})	\\
qr & Power exponent of the temperature distribution (as a function of radius) in the rim \\
qa & Power exponent of the temperature distribution (as a function of radius) in the disk atmosphere \\
qm & Power exponent of the temperature distribution (as a function of radius) in the disk midplane \\
\enddata
\tablecomments{List of parameters involved in the TLTD model, as well as their relevances and implications.
As mentioned in the text, only the mass contributions D$_0$, D$_1$, D$_2$, D$_{i,j}$, and the temperature 
exponents qr, qa, and qm are fitted to the data. }
\end{deluxetable}

\begin{deluxetable}{lcccccccl}
\tabletypesize{\scriptsize}
\rotate
\tablenum{6}
\tablecolumns{9} 
\tablewidth{0pc} 
\tablecaption{Spectral types, extinction, and optical spectral features\label{lines-table}} 
\tablehead{
 \colhead{Name} & \colhead{Sp.Type} &  \colhead{A$_V$} & \colhead{H$\alpha$ (6563\,\AA)}  & \colhead{Li I (6708\,\AA)} & \colhead{K I (7665-7699\,\AA)} & \colhead{Na I (8183-8195\,\AA)$^a$}& \colhead{Na I (8183-8195\,\AA)$^b$} & \colhead{Notes}} 
\startdata
CrA-453 & M4.5  & 1.84$\pm$0.17  & -3.9  & 0.5 & 1.8---1.3 & 1.0---1.5  & ...---1.3 & WTTS, Class III  \\
CrA-465 & M7.5  &  0.08$\pm$0.04 & -253  & 0.5 & 2.7---2.5 & 1.0---1.0  & 1.0---1.0 & CTTS, H$\alpha$ double peak, Class II \\
CrA-466 & M2.0 & 8.10$\pm$0.44  & -14.5 & 0.5 & 1.4---0.8 & 0.9---1.1 & 0.8---1.0   & CTTS, TO \\
CrA-4110 & M5 & 0.41$\pm$0.34 & -10.9 & 0.5 & 1.7---1.4 & 0.9---2.5: & 1.0---1.2 & WTTS, Class II \\
CrA-4111 & M4.5 &  0.00$\pm$0.46  & -12.3 & 0.5 & 2.3---1.9 & 1.2---1.5 & ...---1.5  &  WTTS \\
G-14 & M4.5 & 1.89$\pm$0.10  & -7.4 & 0.6 & 2.1---1.5 & 1.2---1.6 & ...---1.4   & WTTS,TO \\
G-17 & ---	 & ---  & -39	 & ...	 & ...  & ...  & ... & Faint   \\
G-28 & ---	 & ---  & -200	 & ...	 & ...  & ...  & ... &  Faint, Class I:   \\
G-30 & M3.5 & 0.09$\pm$0.50  & -7.5 & 0.3 & 1.8---1.3 & 1.1---1.5 &...---1.4 &  WTTS, Class III \\
G-49 & M4.0 &  0.07$\pm$0.02 & -4.3 & 0.6 & 1.6---1.1 & 1.0---1.3 & 1.0---1.2   & WTTS, Class III  \\
G-64 & late M:$^c$  & --- & -26	 & 1.5:	 & 0.5:---...  & ... & ...  & Class I + HH  \\
G-65 & M1-M2:$^c$  & 14$\pm$1 & -6	 & 0.4:	 & 0.5:---0.2  & 0.5---0.7 & 0.6---0.4  & CTTS (broad H$\alpha$), Class II  \\
G-76 & ---	 & --- & -6	 & ...	 & ...  & ...  & ... &  Faint   \\
G-80 & ---  & --- & -120	 & ...	 & ...  & ...  & ... & Shock, NII,SII  \\
G-85 & M2-M3:$^c$  & 17$\pm$1 & -27	 & ...	 & 2.4:---0.9 & 0.9---1.3  & ... & CTTS, Class II  \\
G-87 & M3-M4:$^c$  & 14$\pm$1 & -4	 & 2.0:	 & 1.4---1.1  & 0.9---1.0  & 0.9---1.1 & WTTS, TO   \\
G-94 & M3.5 & 0.59$\pm$0.02  & -5.6 & 0.1 & 1.7---1.2 & 1.1---1.4 & ...---1.2   & WTTS, Class III \\
G-95 & M1.0 &  5.01$\pm$0.45 & --- & 0.4 & 1.0---0.3 & 0.4---0.6 & ...---0.5   & WTTS, Class III \\
G-102 & M5.0 & 0.72$\pm$0.04  & -15.1 & 0.6 & 2.6---2.1 & 1.4---2.0 &  ...---1.6  & WTTS, Class III \\
G-108 & late M:$^c$ & --- & ... & ... &  0.8:---0.4:  & ... & ... & High extinction   \\
G-122 & ---   	 & --- & -34	 & ...	 & ...  & ...  & ... & Class I, Faint   \\
G-133 & K-M:$^c$  & ---	 & -4	 & 0.3:	 & ...  & ...  & ... & High extinction, Class I:, H$\alpha$ marginal    \\
\enddata
\tablecomments{Equivalent widths of the lines observed in the young stars. 
EW in \AA, negative values mean emission. $^a$: From the spectra centered at 773 nm.
$^b$: From the spectra centered at 881 nm. $^c$: Spectral type obtained by comparison
with similar type spectra because of low S/N. Colons mark uncertain values. The error in A$_V$ is
a combination of uncertainty in the spectral type and dispersion in the A$_V$ values obtained
from J-H and J-K. The ``Notes'' colum lists the type of object based on the strength and broadness of the
H$\alpha$ line (CTTS if EW$>$10 and/or broad line; WTTS otherwise) and the SED (Class I, Class II, TO,
and Class III).}
\end{deluxetable}

\begin{deluxetable}{lcccccc}
\rotate
\tablenum{7}
\tablecolumns{7} 
\tablewidth{0pc} 
\tablecaption{Silicate composition\label{sil-table}} 
\tablehead{
 \colhead{Name} & \colhead{$\chi ^2$}& \colhead{Am. (Olivine-type)} & \colhead{Am. (Pyroxene-type)} & \colhead{Forsterite} & \colhead{Enstatite} & \colhead{Silica}   \\
		& & \colhead{0.1$\mu$m}   & \colhead{0.1$\mu$m}    & \colhead{0.1$\mu$m}  & \colhead{0.1$\mu$m} & \colhead{0.1$\mu$m} \\
		& & \colhead{1.5$\mu$m}   & \colhead{1.5$\mu$m}    & \colhead{1.5$\mu$m}  & \colhead{1.5$\mu$m} & \colhead{1.5$\mu$m} \\
		& & \colhead{6.0$\mu$m}   & \colhead{6.0$\mu$m}    & \colhead{6.0$\mu$m}  &  		      & \colhead{6.0$\mu$m}     } 
\startdata
CrA-466 & 10.2 & --- & --- & 5.2$^{+0.3}_{-0.3}$ & 0.01$^{+0.24}_{-0.01}$ & ---  \\
 & & 77$^{+3}_{-3}$ & --- & --- & 1.9$^{+1.3}_{-1.1}$ & ---  \\
 & & --- & --- & --- & --- & 16$^{+2}_{-2}$  \\
CrA-466$^l$ & 9.3 & 1$^{+22}_{-1}$ & 0$^{+10}_{-0}$ & 9$^{+10}_{-3}$ & --- & ---  \\
 & & 28$^{+78}_{-26}$ & 4$^{+12}_{-3}$ & --- & --- & 0.4$^{+2.4}_{-0.4}$  \\
 & & --- & 54$^{+38}_{-42}$ & --- & --- & 3.8$^{+8.6}_{-2.4}$  \\
CrA-4107 & 1.3  & 0.2$^{+3.1}_{-0.2}$ & 2$^{+11}_{-2}$ & 2.0$^{+1.1}_{-0.7}$ & 1.2$^{+1.7}_{-0.8}$ & 0.11$^{+0.61}_{-0.11}$  \\
 & & 2.6$^{+7.4}_{-2.5}$ & 0.7$^{+6.0}_{-0.7}$ & 0.0$^{+2.0}_{-0.0}$ & 2.2$^{+2.6}_{-1.6}$ & 1.3$^{+1.3}_{-0.9}$  \\
 & & 73$^{+25}_{-42}$ & 3$^{+25}_{-3}$ & 11$^{+12}_{-8}$ & --- & 1.3$^{+5.0}_{-1.3}$  \\
CrA-4111$^l$ & 1.9 & --- & --- & 11$^{+5}_{-5.}$ & 3$^{+47}_{-3}$ & ---  \\
 & & --- & --- & --- & 62$^{+13}_{-18}$ & 7$^{+13}_{-66}$  \\
 & & --- & --- & --- & --- & 18$^{+21}_{-17}$  \\
G-1 & 6.2 & 2$^{+13}_{-2}$ & 0.08$^{+0.84}_{-0.08}$ & 0.48$^{+0.23}_{-0.27}$ & 0.02$^{+0.14}_{-0.02}$ & 0.56$^{+0.16}_{-0.31}$  \\
 & & --- & 13.6$^{+2.9}_{-9.2}$ & 0.12$^{+0.27}_{-0.12}$ & 2.1$^{+0.8}_{-1.2}$ & 0.4$^{+0.3}_{-0.3}$  \\
 & & 6$^{+17}_{-6}$ & 73$^{+8}_{-26}$ & 0.4$^{+2.9}_{-0.4}$ & --- & 1.2$^{+1.8}_{-1.1}$  \\
G-14 & 6.5 & 0.7$^{+2.7}_{-0.7}$ & --- & 2.7$^{+0.2}_{-0.2}$ & --- & ---  \\
 & & 32$^{+2}_{-2}$ & --- & 0.05$^{+0.56}_{-0.05}$ & --- & ---  \\
 & & --- & 64$^{+2}_{-4}$ & 0.02$^{+1.67}_{-0.02}$ & --- & ---  \\
G-85 & 6.1  & 11$^{+2}_{-1}$ & --- & 3.4$^{+0.5}_{-0.3}$ & 2.9$^{+0.4}_{-0.3}$ & ---  \\
 & & --- & 61$^{+5}_{-4}$ & --- & 2.4$^{+0.8}_{-0.8}$ & 0.36$^{+0.30}_{-0.29}$  \\
 & & 0.4$^{+4.6}_{-0.4}$ & 14$^{+10}_{-10}$ & --- & --- & 4.6$^{+1.6}_{-1.8}$  \\
G-85$^l$ & 1.1 & 2$^{+19}_{-2}$ & 26$^{+29}_{-18}$ & 9.5$^{+7.1}_{-2.6}$ & 0.8$^{+3.8}_{-0.8}$ & 0.02$^{+0.97}_{-0.02}$  \\
 & & 53$^{+19}_{-34}$ & 3$^{+21}_{-3}$ & --- & 5$^{+3}_{-3}$ & 0.36$^{+0.93}_{-0.34}$  \\
 & & --- & --- & --- & --- & 0.3$^{+5.1}_{-0.3}$  \\
G-87 & 3.4 & 0.3$^{+5.6}_{-0.3}$ & --- & 12$^{+3}_{-2}$ & 11$^{+64}_{-4}$ & ---  \\
 & & 55$^{+16}_{-35}$ & 19$^{+47}_{-18}$ & 0.3$^{+5.0}_{-0.3}$ & 0.2$^{+3.2}_{-0.2}$ & ---  \\
 & & --- & --- & 0.6$^{+11.4}_{-0.6}$ & --- & 2$^{+2}_{-2}$  \\
\enddata
\tablecomments{Percentage of the mass fraction of the different materials and
reduced $\chi ^2$ for each fitted spectrum. The amorphous
silicates are classified as having olivine or pyroxene stoichiometry. The crystalline
silicates are forsterite, enstatite, and silica. Amorphous carbon grains are
included as well. For each component, three different sizes are given (0.1, 1.5 and 6.0 $\mu$m).
The summary lists the final crystallinity fraction and the mass average grain size (including
crystalline and amorphous grains). $^l$= Long wavelength composition (17-25$\mu$m).}
\end{deluxetable}

\clearpage

\begin{deluxetable}{lccc}
\tabletypesize{\scriptsize}
\tablenum{8}
\tablecolumns{4} 
\tablewidth{0pc} 
\tablecaption{Grain sizes and crystallinity fraction \label{sizecryst-table}} 
\tablehead{
 \colhead{Name}  & \colhead{Size (am.) / $\mu$m} & \colhead{Size (cryst.) / $\mu$m} & \colhead{Cryst. (\%)}} 
\startdata
CrA-466 & 1.5$\pm$0.1 &  4.3$\pm$0.5 & 23$\pm$2  \\
CrA-466$^l$ & 4.3$\pm$2.9 &  1.8$\pm$2.1 & 14$\pm$7  \\
CrA-4107 & 5.7$\pm$2.6 &  4.1$\pm$3.3 & 19$\pm$11  \\
CrA-4111$^l$ & --- &  2.1$\pm$1.2 & 70-100$^*$  \\
G-1 & 5.2$\pm$1.2 &  2.6$\pm$2.1 & 5$\pm$2  \\
G-14 & 4.5$\pm$0.2 &  0.2$\pm$0.4 & 2.8$\pm$0.3  \\
G-85 & 2.1$\pm$0.7 &  2.4$\pm$0.8 & 14$\pm$2  \\
G-85$^l$ & 1.0$\pm$0.5 &  0.7$\pm$0.6 & 16$\pm$6  \\
G-87 & 1.5$\pm$0.8 &  0.7$\pm$1.2 & 26$\pm$8  \\
\enddata
\tablecomments{Average grain sizes for amorphous and crystalline silicates 
and crystallinity fraction (mass fraction of crystalline silicates) from the TLTD fit.
$*$ For the low S/N spectrum of CrA-4111$^l$, there is only marginal evidence of 
crystalline silicates, so the crystallinity fraction is constrained taking into 
account the errors in the marginally detected crystalline silicates. The typical error for
this value is $\sim$30\%.
}
\end{deluxetable}

\begin{deluxetable}{lccccc}
\tabletypesize{\scriptsize}
\tablenum{9}
\tablecolumns{5} 
\tablewidth{0pc} 
\tablecaption{Excesses over the photosphere for the disks with inner holes \label{excess-table}} 
\tablehead{
 \colhead{Name}  & \colhead{log(F$_{10}$/F$_{10*}$)}& \colhead{log(F$_{24}$/F$_{24*}$)}& \colhead{log(F$_{30}$/F$_{30*}$)}& \colhead{log(F$_{70}$/F$_{70*}$)} & \colhead{Disk Type} } 
\startdata
CrA-205  & 0.0 & 1.3 & 1.3: & --- & TO \\
CrA-4109 & 0.0 & 1.5 & 1.8  & --- & TO \\
CrA-4111 & 0.0 & 1.9: & 2.0: & --- & TO \\
G-1      & 0.7 & 1.3 & 1.6: & 1.9  & TO \\
G-30     & --- & 0.7 & --- & --- & D \\
G-87     & 0.6 & 1.5 & 1.6 & --- & TO \\
G-94     & 0.0 & 0.4: & --- &  --- & D \\
G-95     & 0.0 & 0.5  & 0.8 &  --- & D \\
G-102    & --- & 0.5 & --- &  --- & D \\
\enddata
\tablecomments{Excesses over the photospheric levels for the disks with inner holes.
Colon (:) indicates uncertain values. Otherwise, the typical errors are $\pm$0.1.
The disk type is marked as ``transition'' (TO) or ``debris-like'' (D) disk depending on
the excess over the photosphere (Kenyon \& Bromley 2005). Note that the debris-like
disks are considered to be Class III sources and not included in the median SED.}
\end{deluxetable}

\begin{deluxetable}{lcccc}
\tablenum{10}
\tablecolumns{5} 
\tablewidth{0pc} 
\tablecaption{Forbidden lines and shocks \label{shock-table}} 
\tablehead{
 \colhead{Name} & \colhead{[N II] 6548\AA} &  \colhead{[N II] 6583\AA} & \colhead{[S II] 6716\AA}  & \colhead{[S II] 6731\AA} } 
\startdata
G-64 & -2.6 & -2.7 & -19.2 & -22.8 \\
G-80 & -18.1 & -68.1 & -48.3 & -43.0 \\
\enddata
\tablecomments{Equivalent widths of the forbidden lines observed in the HH objects 
detected in the spectra.}
\end{deluxetable}

\clearpage

\clearpage

\clearpage

\clearpage

\clearpage

\clearpage

\clearpage

\clearpage

\clearpage

\clearpage

\clearpage

\clearpage

\clearpage

\end{document}